\documentclass{aa2}
\usepackage[dvips]{graphicx}

\usepackage{epsfig}
\usepackage{natbib}
\bibpunct{(}{)}{;}{a}{}{,}

\begin{document}
\title{{\it HIPPARCOS} age-metallicity relation of the solar neighbourhood disc stars}
\author{ A. Ibukiyama \inst{1} \and N. Arimoto \inst{1,2}}
\offprints{A. Ibukiyama}
\mail{aibukiya@optik.mtk.nao.ac.jp}

\institute {Institute of Astronomy (IoA), School of Science,
        University of Tokyo, 2-21-1 Osawa, Mitaka, Tokyo 181-0015, Japan
	\and National Astronomical Observatory, 2-21-1 Osawa, Mitaka, Tokyo 181-8588, Japan}

\date {Received  /Accepted 1 July 2002}
\titlerunning {{\it HIPPARCOS} age-metallicity relation}
\authorrunning {A. Ibukiyama \& N. Arimoto}

\abstract{
We derive age-metallicity relations (AMRs) and orbital parameters for the 1658 solar neighbourhood stars to which accurate distances are measured by the {\it HIPPARCOS} satellite. 
The sample stars comprise 1382 thin disc stars, 229 thick disc stars, and 47 halo stars according to their orbital parameters.
We find a considerable scatter for thin disc AMR along the one-zone Galactic chemical evolution (GCE) model. 
Orbits and metallicities of thin disc stars show now clear relation each other. The scatter along the AMR exists even if the stars with the same orbits are selected. 
We examine simple extension of one-zone GCE models which account for inhomogeneity in the effective yield and inhomogeneous star formation rate in the Galaxy. 
Both  extensions of one-zone GCE model cannot account for the scatter in age - [Fe/H] - [Ca/Fe] relation simultaneously. We conclude, therefore, that the scatter along the thin disc AMR is an essential feature in the formation and evolution of the Galaxy.
The AMR for thick disc stars shows that the star formation terminated 8 Gyr ago in thick disc. 
As already reported by \citet{Gratton_et.al.2000} and \citet{Prochaska_et.al.2000}, thick disc stars are more Ca-rich than thin disc stars with the same [Fe/H].
We find that thick disc stars show a vertical abundance gradient. 
These three facts, the AMR, vertical gradient, and [Ca/Fe]-[Fe/H] relation, support monolithic collapse and/or accretion of satellite dwarf galaxy as thick disc formation scenario.
}
	
\maketitle

\section{Introduction}
The individual ages for solar neighbourhood stars are indispensable in the research of star formation history of the Galaxy. \citet{Twarog1980} first derived 
the age-metallicity relation (AMR) for the disc in the neighbourhood of the Sun from
{\it ubvy} and H$\beta$ photometry of a large
sample of field stars. Theoretical isochrones used
in the age determination were taken from \citet{Ciardullo_Demarque1977}.
In the Twarog's AMR, the metallicity increases from 
[Fe/H]$=-1.0$ at 13 Gyr to [Fe/H]$=-0.03$ at the age of the Sun.
The mean metallicity has increased more slowly since then
to a present value of [Fe/H]$=+0.01$ for the youngest stars.
The dispersion in [Fe/H] is as small as $\pm 0.1$ dex 
at any given age. \citet{Carlberg_et.al.1985} 
used stellar models of \citet{VandenBerg1983}
and a revised metallicity calibration
that takes into account a temperature dependence. 
Both ages and metallicities were estimated in a photometric
manner, and the resulting AMR is qualitatively similar
to that of \citet{Twarog1980}, but [Fe/H] increases more
gradually, showing an increase of only 0.3 dex over the past
15 Gyrs \citep[cf.][]{Nissen_Schuster1991}.
The metallicity dispersion decreases from 
$\pm 0.15$ dex for the oldest stars 
($13-20$ Gyrs) to $\pm 0.05$ dex for younger stars. 

\citet{Edvardsson_et.al.1993} derived elemental abundances
of O, Na, Mg, Al, Si, Ca, Ti, Fe, Ni, Y, Zr, Ba, and Nd
for 189 nearby long-lived disc dwarfs by using 
high resolution, high S/N, spectroscopic data. Individual
ages were derived photometrically from fits in the 
$\log T_{\rm eff} - \log g$ plane of 
the isochrones by \citet{VandenBerg1985}. 
The uncertainties in the relative ages are about
$25 \%$. Due to metallicity measurements of high 
precision, \citet{Edvardsson_et.al.1993}
improved greatly the AMR, but ironically the resulting AMR
clearly indicated a considerable 
scatter ($\sim 0.15$ dex)
in the metallicities of disc stars formed at 
any given time, implying that there is only a very weak
correlation between age and metallicity. The scatter
seems to be substantially larger than that can be explained 
by observational errors. If the scatter is real, it would 
cause a serious difficulty for galactic chemical evolution (GCE) models, because 
it is easy to fit the average run of the data, but 
difficult to explain such a large scatter without breaking
some of assumptions that GCE models usually make \citep{Pagel_Tautvaisiene1995}. 
\citet{Edvardsson_et.al.1993} suggested that the scatter arises
from star formation stimulated from sporadic episodes
of gas infall, although it is also possible that a 
different rate of chemical enrichment, depending 
on the distance from the Galactic centre, causes a 
scatter of this kind.

In this article, we derived the ages and orbital parameters for 1658 solar neighbourhood stars, almost ten times more than the previous researches.  Hence we succeed in finding out 
the new features of the Galaxy using stellar ages, chemical components, and orbits.
The article is organised as follows. Section 2 derives the ages and the orbital parameters for sample stars. Section 3 shows the features relevant to thin disc stars while Sect. 4 shows those of thick disc. Section 5 discusses observational error in our data, abundance gradient, the abundance distribution functions, the formation of thick disc, and the scatter along the thin disc AMR. Section 6 concludes the present study.

\section{The Data}
\subsection{The observational data}
\subsubsection {Absolute magnitudes and colours}
Visual magnitudes, $B-V$ colours, and parallaxes of nearby field stars
were all taken from the {\it HIPPARCOS} catalogue \citep{hipparcos}.
The absolute magnitudes, $M_{V_0}$, and unreddened colours, $(B-V)_0$, 
were then calculated by applying the reddening corrections given by the model of \citet{Arenou_et.al.1992}.
%We obtained 2123 stars by cross-checking the {\it HIPPARCOS} catalogue 
%and the [Fe/H] catalogue of Cayrel de Strobel et al. (1998). 
Certain number of stars in the {\it HIPPARCOS} catalogue suffer 
from uncertainties in parallaxes, binarities, and variabilities. 
Thus, the following criteria were introduced in our sample selection: 
(1) $\sigma_{\pi}/{\pi} <0.1 $, where $\pi$ and 
$\sigma_{\pi}$ are the parallax and the dispersion, respectively; 
(2) no binaries -- binaries were excluded if explicitly identified 
in the {\it HIPPARCOS} catalogue; 
(3) non-variables -- variables were excluded 
if assigned in the {\it HIPPARCOS} catalogue or in the [Fe/H] catalogue of \citet{CayrelCatalogue};
(4) no giants -- giants were excluded according to their location on the CM diagram;
%(6) no field horizontal branch (HB) stars -- HB stars were excluded according to 
%Gratton (1998),  \citet{Altmann_DeBoer2000},
%and Wilhelm et al. (1999); 
%(7) no $\lambda$ Bootis stars -- $\lambda$ Bootis stars are chemically 
%peculiar stars showing low iron abundance (typically 
%$[Fe/H] \simeq -1.5$), although their oxygen and sulphur abundances are 
%similar to those of the Sun. $\lambda$ Bootis stars were excluded according to 
%Faraggiana \& Bonifacio (1999). 
%From 2123 whole sample, 765 stars of poor
%parallax data, 315 stars of low quality photometry,
%714 binaries, 381 variables, 1087 giants, 13 HB stars, 15 $\lambda$ 
%Bootis stars, and 244 stars observed before 1980 were excluded because of their relatively lower
%quality of [Fe/H] data. In total, 366 stars remain as our sample.
                        %%%% Fig. 1 %%%%
                        %%%% Fig. 2 %%%%
                        %%%% Table  1 %%%%
%Table 1 gives $M_{\rm V_0}$, $(B-V)_0$, $\pi$, $\sigma_{\pi}$, 
%and $E(B-V)$ of the sample stars. 
LK bias \citep{Lutz_Kelker1973} was not 
corrected because \citet{Carretta_et.al.1999} has shown recently that the bias is negligible if a  
strict sample selection ($\sigma_\pi/\pi < 0.12 $) is
applied \citep[see also][]{Ng_Bertelli1998}.

\subsubsection {Metallicities}
The stellar metallicities for F, G, and K stars were taken from the latest [Fe/H]
catalogue of \citet{CayrelCatalogue} which compiles the 
[Fe/H] values derived from high S/N and high resolution spectra. 
We cross-identified 
the {\it HIPPARCOS} catalogue and the [Fe/H] catalogue
to derive a set of reliable
$M_{\rm V_0}$, $(B-V)_0$, and [Fe/H] for 429 stars.
When several [Fe/H] values were given in the [Fe/H] catalogue, 
we adopted the newest observation. Since we wish
to use preferentially the CCD data, we excluded the [Fe/H] data before 1985. 

Additionally we took $ubvy$-H$\beta$ photometry for F and G stars from the 
catalogue of \citet{Hauck_Mermilliod1998}. We calculated [Fe/H] 
from these data adopting the calibration by \citet{Schuster_Nissen1989}.
We define the AMR using spectroscopic metallicity of \citet{CayrelCatalogue}
as ``spectroscopic AMR'' and the AMR using photometric metallicity derived from the 
data of \citet{Hauck_Mermilliod1998} as ``photometric AMR'' hereafter.

Abundances of Ca were taken from \citet{Boesgaard_Tripicco1986}, \citet{Smith_Lambert1986}, \citet{Smith_Lambert1987}, \citet{Hartmann_Gehren1988}, \citet{Abia_et.al.1988}, \citet{Magain1989}, \citet{Cayrel_Bentolila1989}, \citet{Gratton_Sneden1991}, \citet{Berthet1991}, \citet{Edvardsson_et.al.1993}, \citet{Pilachowski_et.al.1993}, \citet{Smith_et.al.1993}, \citet{Nissen_et.al.1994}, \citet{Beveridge_Sneden1994}, \citet{Nissen_Schuster1997}, \citet{Tomkin_et.al.1997}, \citet{Carney_et.al.1997}, \citet{Castro_et.al.1997}, \citet{Giridhar_et.al.1997}, \citet{Gonzalez1998}, \citet{Feltzing_Gustafsson1998}, \citet{King_et.al.1998}, \citet{Jehin_et.al.1999}, \citet{Clementini_et.al.1999}, \citet{Sadakane_et.al.1999}, \citet{Santos_et.al.2000}, \citet{Thoren_Feltzing2000}, \citet{Chen_et.al.2000}, \citet{Gonzalez_Laws2000}, and \citet{Fulbright2000}. Ca abundances may be less accurate than Fe abundances, for the number of Ca lines used in data reduction is fewer than that of Fe.

\subsubsection{The kinematic data}

We adopted radial velocities from the {\it HIPPARCOS} Input Catalogue \citep{hic}, \citet{Barbier_et.al.1994}, \citet{web}, and \citet{Malaroda_et.al.2001}. 
Proper motions were taken from the {\it HIPPARCOS} Catalogue.
The space motions relative to the Sun, $(U,V,W)$, were calculated using \citet{Johnson_Soderblom1987,Edvardsson_et.al.1993}.

We obtained 4240 sample stars by cross checking {\it HIPPARCOS} data, [Fe/H] or $ubvy$-H$\beta$ catalogue, and radial velocity catalogue. 
From 4240 whole sample, 1380 stars of poor
parallax data, 1311 binaries,  428 variables, 1252 giants, 
and 133 stars observed before 1985 were excluded.
 In total, 1658 stars remained as our sample including 489 stars with spectroscopic [Fe/H] and 1169 stars with photometric [Fe/H]. [Ca/Fe] values were obtained for 277 stars among them. 

\begin{table*}
\caption{The sample stars.}
\begin{tabular}{lrrrr}
\hline\hline
& thin disc & thick disc & halo & total \\
\hline
photometric [Fe/H] & 1010 & 143 & 16 &  1169 \\
spectroscopic [Fe/H] (without [Ca/Fe] ) & 173 &28 &11 & 212 \\
spectroscopic [Fe/H] (with [Ca/Fe])  & 199 & 58 & 20& 277 \\
total & 1382 & 229 & 47 & 1658\\
\hline
\end{tabular}
\label{sample_table}
\end{table*}
	
\begin{figure*}
\includegraphics[width=17cm,height=11cm]{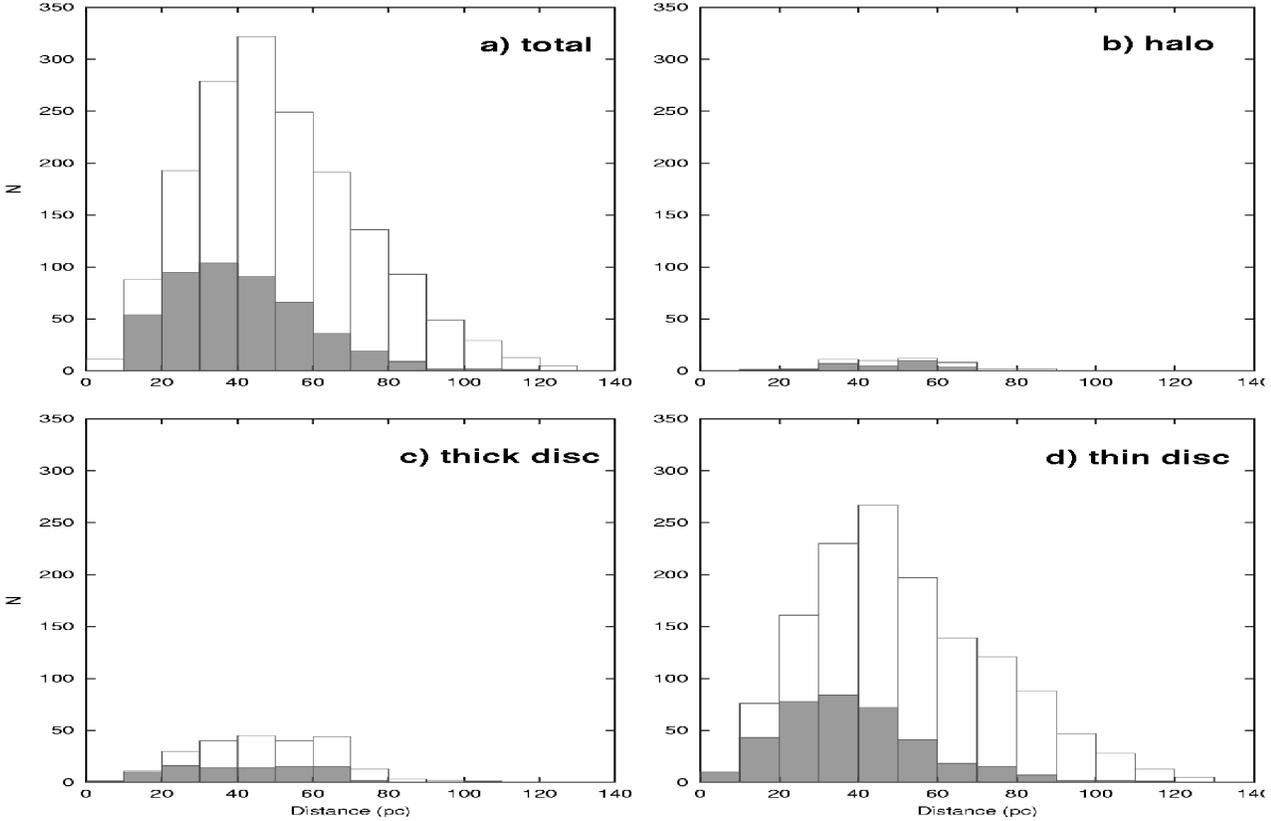}
\caption{The distribution of distances of the sample stars from the sun. \textbf{a)}  whole 1658 stars. \textbf{b)} 47 halo stars. \textbf{c)} 229 thick disc stars. \textbf{d)} 1382 thin disc stars. Gray boxes show the stars with spectroscopic data while open boxes represent the stars with photometric data.}
\label{distance_distribution}
\end{figure*}

Figure \ref{distance_distribution} shows the distances from the Sun of the sample stars.
The sample stars are currently locating at the distances 
from 10 pc to 120 pc with a median value of 40 pc.

%%%%%%%%%%%%%%%% Fig.  %%%%%%%%%%%%%%%%%

\subsection{The Analysis}
\subsubsection {Ages}

 Once $M_{\rm V_0}$, $(B-V)_0$, and [Fe/H] were known, it was direct 
to derive ages of the selected stars by using the 
isochrone fitting. We adopted Yonsei-Yale isochrones 
\citep{Yi_et.al.2001}, which were calculated with new OPAL opacities 
and Kurucz model atmospheres for a set of metallicities $Z=$ 0.00001, 0.0001, 0.0004, 0.001, 0.004, 0.007, 0.01, 0.02 (solar), 0.04, 0.06, 0.08 and ages 
from $t=$1.0 Gyr to 20.0 Gyr with an interval of $\Delta t=$1.0 Gyr. The mixing length
and helium enrichment rate were fixed to $\alpha\equiv l/H_p=1.7$ 
and $\Delta Y/\Delta Z=2$, respectively, while no convective overshooting was 
introduced. Since Yonsei-Yale isochrones were calculated for
$Z$ instead of [Fe/H], we converted [Fe/H] of the selected 
stars into $Z$ by using an empirical relation
for the solar neighbourhood stars \citep{Clementini_et.al.1999}:
\begin{eqnarray}
\log(Z/Z_{\odot})\simeq {\rm [\alpha/H]} & \nonumber\\
& \hspace{-0.5cm} \simeq \left\{
\begin{array}{ll}
{\rm [Fe/H]}+0.4, &{\rm [Fe/H]} \le  -1.0\cr
0.6{\rm [Fe/H]}, & {\rm [Fe/H]} \ge -1.0\\
\end{array}
\right.
\end{eqnarray}
where $Z_{\odot}=0.02$, and $\alpha$ elements are assumed to dominate $Z$.
Admittedly, this is rather a crude approximation, but it would
not introduce significant effects on the resulting AMR and, 
we believe, is far better than a simple assumption 
of $\log(Z/Z_{\odot})=$[Fe/H]. We also calculated ages of 
the sample stars with the assumption of $\log(Z/Z_{\odot})=$[Fe/H] to find that the
resulting ages are systematically older by $\sim 0.2$ dex for
metal-poor stars ([Fe/H]$ \le -1$). The difference becomes smaller
toward higher [Fe/H].

\begin{figure}
\resizebox{\hsize}{!}{\includegraphics{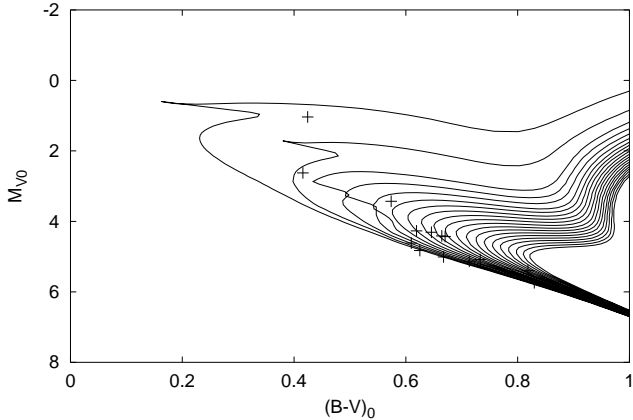}}
\caption{The sample of isochrone fitting. Crosses represent solar metallicity stars ($0.0 \leq $[Fe/H]$ <0.005$) while solid lines represent the isochrones for solar metallicity stars from 1 Gyr (top left) to 20 Gyr (bottom right).}
\label{iso}
\end{figure}

Finally, knowing $Z$ for the star, we interpolated the isochrones 
in $\log Z$ with the two adjacent metallicities and derived a set
of the isochrones with $Z$ and ages from 1.0 to 20.0 Gyr. As illustrated
in Fig. \ref{iso}, the stellar age was derived by interpolating linearly
the nearest two grid points of the isochrones on the $M_{\rm V_0} - (B-V)_0$ diagram. 
%The resulting ages for the 366 sample
%stars are given in Table 2. We used the metallicity based on $\log{Z/Z_{\odot}}=[Fe/H] $ 
%for columns (1) to (4), while Eq. (1) (oxygen enhanced model) was adopted for columns (5) to (8). All the ages are given in units of log Gyr. 
Uncertainties in $M_{V_0}$, $(B-V)_0$, and $Z$ cause errors for resulting ages. 
Errors in parallax of 10\%, $(B-V)_0$ of 0.02 mag, and metallicity of 0.1 dex result in uncertainties in age of 0.064 dex, 0.073 dex, 0.082 dex, respectively. 
Thus 0.12 dex age error exists for our sample in average.

%%%%%%%%%%%%% Fig.  %%%%%%%%%%%%%%%%

\subsubsection{The Galactic orbital parameters}
We examined kinematics of the sample stars and 
calculated their orbital parameters in a similar way 
to \citet{Edvardsson_et.al.1993} and \citet{Nissen_Schuster1997}.

We integrated their orbit backward in time by using a four-component Miyamoto-Nagai potential 
model including thin disc, thick disc, bulge, and halo \citep{Sofue1996}. 
Integration was done using 6th order Symplectic formula 
with a constant time step of $10^5$ year. 
We adopted the correction for the solar velocity of $(U_{\odot},V_{\odot},W_{\odot}) = (-10,232,6) \mbox{ km s}^{-1}$. 
The galactocentric distance and circular velocity of the sun were fixed to be 
8.0 kpc and 226.0 $\mbox{ km s}^{-1}$, respectively \citep{Edvardsson_et.al.1993}.
Energy and angular momentum of the sample stars were 
conserved within an accuracy of $10^{-8}$. 
We derived apo-galactocentric distance projected to the Galactic plane, $R_a$, peri-galactocentric  distance, $R_p$, 
and maximum deviation from the plane, $z_{\rm max} $. We also derived eccentricity, $e \equiv (R_p-R_a) / (R_p + R_a) $, and mean distance, $R \equiv (R_p+R_a) /2$. Tables \ref{spec_table} \& \ref{photo_table} give the orbital parameters of the sample stars. 

%Then we integrated the orbit backward to the age of the star concerned, of which location in that time  is defined as their ``birth place''. 
%Fig. 4 show the birth places of the sample stars. Although the birth places suffer from the large uncertainty in the age, their total positions shows their feature well because of the orbit in such a potential is restricted into a torus. 

%\section{Stellar population in the solar torus}

%The solar neighbourhood is traditionally considered as a closed region of the small radius (in this case, 100pc).
%The spatial distribution of 366 {\it HIPPARCOS} stars in the ``100pc sphere'' is only a snap shot of the solar neighbourhood at the present epoch. 
%The sample stars can be born in the different places.
%We integrated orbits of individual stars backward to their ages, 
%and defined the locations at that time as the ``birth place'' of the stars.
%We found that the birth place of the stars is confined in a torus of which $R_p=6\mbox{ kpc}$, $R_a=9\mbox{ kpc}$, and $|z_{\rm max}|
%=0.3\mbox{ kpc}$ (Fig. 4).  
%Although precise locations of the birth place is rather difficult to determine 
%due to large uncertainties in age,
%global features should not change much because the orbits in such a potential should be restricted 
%within a torus (Binney \& Tremaine 1987). Hereafter, we define this region as the solar torus in which the ``solar neighbourhood stars'' formed in the past. 
%In the solar torus, the sample stars are composed of three populations; namely, thin disc, thick disc, and halo. 
We identified the populations of stars by using their rotational velocities, $V$, 
with respect to the Galaxy centre. 
The stars with $V \ge  -62 \mbox{ km s}^{-1}$, 
$-182 \mbox{ km s}^{-1} \le V < -62 \mbox{ km s}^{-1}$,
and $ V < -182 \mbox{ km s}^{-1}$ are assigned as thin disc, thick disc, and halo stars, respectively.  
Following \citet{Carney_et.al.1989} and \citet{Prochaska_et.al.2000}, stars with $z_{\rm max} > $ 600 pc are identified to be thick disc component even if their rotational velocities, $V$, are larger than 170 $\mbox{ km s}^{-1}$ in addition. 

In our 1658 sample stars, we identified 1138 thin disc stars, 229 thick disc stars, and 47 halo stars. 

\begin{figure*}
\centering
\includegraphics[width=17cm,height=14cm]{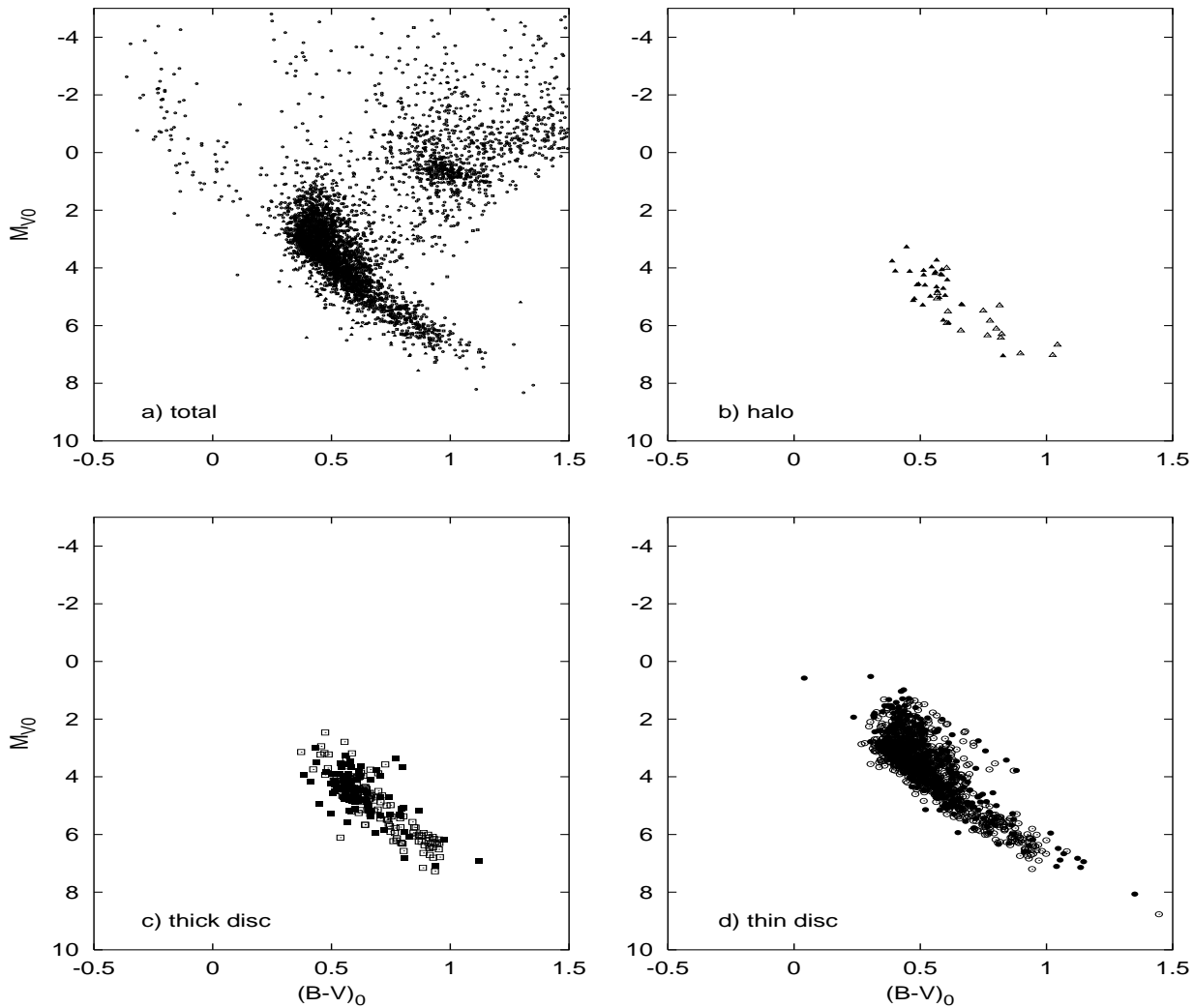}
\caption{Colour-magnitude diagram for the sample stars. \textbf{a)} 4240 stars before sample selection. \textbf{b)} \textbf{c)}, and \textbf{d)} the selected samples. Filled symbols represent the star with spectroscopic data, while open symbols represent stars with photometric data. Circles, squares and triangles represent thin disc, thick disc and halo stars, respectively.}
\label{CM_diagram}
\end{figure*}

The identified stellar populations are listed in Tables \ref{spec_table} \& \ref{photo_table}. 
Figures \ref{CM_diagram} show the CM diagrams of the whole sample with [Fe/H] measurements before sample selection (4240 stars), the selected 1138 thin disc stars, 229 thick disc stars, and 47 halo stars, respectively.

%%%%%%%%%%%%%%%%%%%%% figure %%%%%%%%%%%%%%%%%%%%%

\begin{figure*}
\centering
\includegraphics[width=17cm,height=14cm]{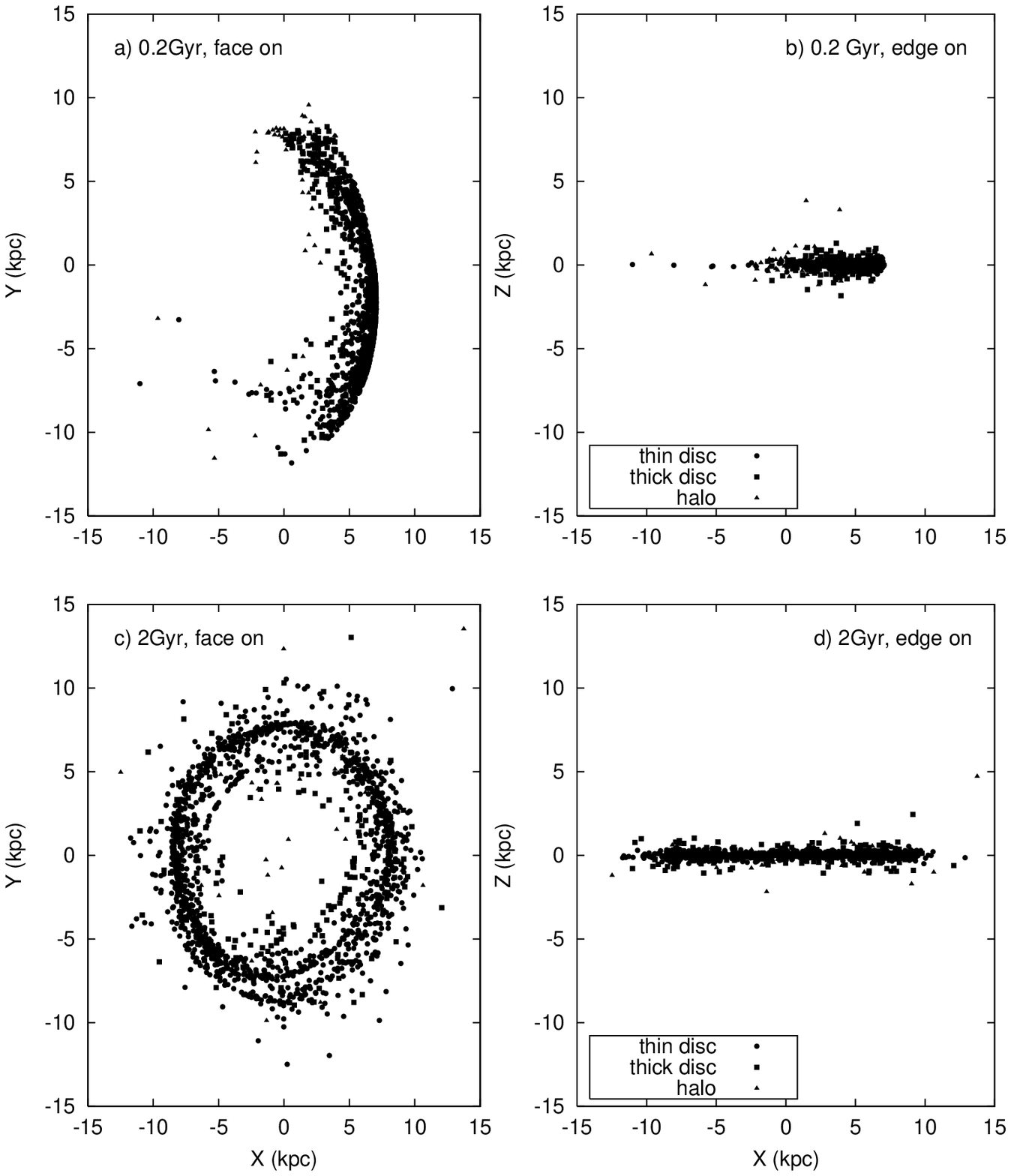}
\caption{Snapshot of the orbits of our sample stars. \textbf{a)} and \textbf{b)} the positions of stars of 0.2 Gyr ago. \textbf{c)} and \textbf{d)} these of 2 Gyr ago. The left panels are face on views and the right panels are edge on views.}
\label{snap}
\end{figure*}

Thin disc stars rotate, in average, around the Galactic centre every 0.2 Gyr. 
For example, a 10 Gyr old star has rotated 50 times till now.
Figures \ref{snap} show the snapshots of the orbit of our sample stars when we integrate their orbit backward to 0.2 Gyr and 2 Gyr. 
In the top two figures, the majority of stars rotate almost once. In such a short time scale, the orbits did not deviate so much and the similarity preserved. 
In the bottom figures, in which the orbits were integrated for 2 Gyr, however, the stars rotate for so long times that they are uniformly distributed in the torus because of their random motion. 
As a result, the sample stars older than 2 Gyr trace the star formation history 
in the various region in the torus, in spite of the fact that 
they now exist in the region within 100 pc from the sun.
Hereafter, we define this region as the solar torus in which the ``solar neighbourhood stars'' formed in the past. 
Our sample stars reflect the star formation history not in the narrow sphere with radius of 100 pc but in the wide torus region of which radius is 6 kpc $\sim$ 9 kpc from the Galactic centre described in Figs \ref{snap}.

%%%%%%%%%% figure %%%%%%%%%%%%%%5

\section{The Thin Disc}
\subsection{The thin disc AMR}

\begin{figure*}
\sidecaption
\includegraphics[width=12cm]{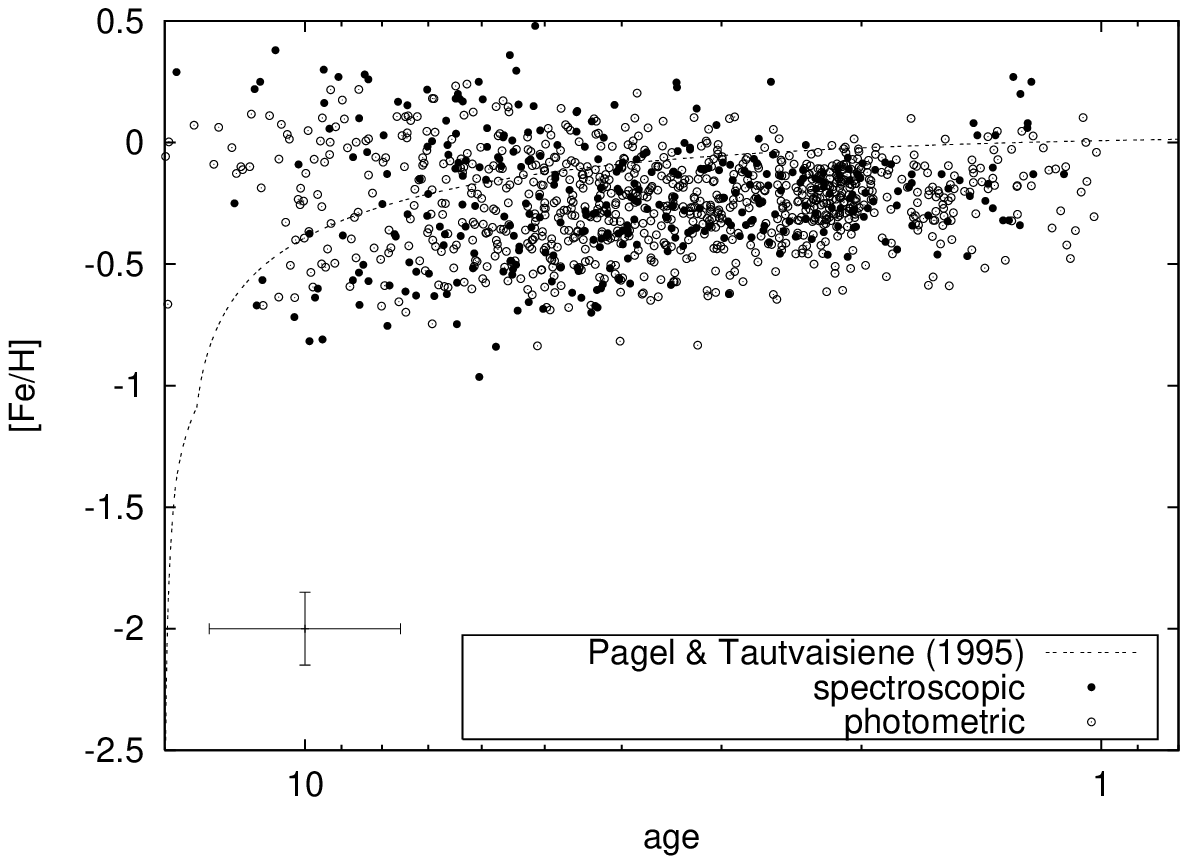}
\caption{The spectroscopic and photometric AMR for thin disc stars.}
\label{thin_AMR_spec}
\end{figure*}

Figure \ref{thin_AMR_spec} shows the spectroscopic AMR and the photometric AMR of thin disc stars. 
The line represents the GCE model by \citet{Pagel_Tautvaisiene1995}, which includes delayed elements production by type Ia supernovae (SNIa) and low mass type II supernovae (SNII).
Thin disc AMR has a considerable scatter.
The scatter is larger than that expected from observational errors.
Thin disc star seems to appear at the beginning of the Galaxy formation although
we are not sure to this feature because such stars have large errors in age determination due to heavy crowding of main sequence isochrones.

\begin{figure}
\resizebox{\hsize}{!}{\includegraphics{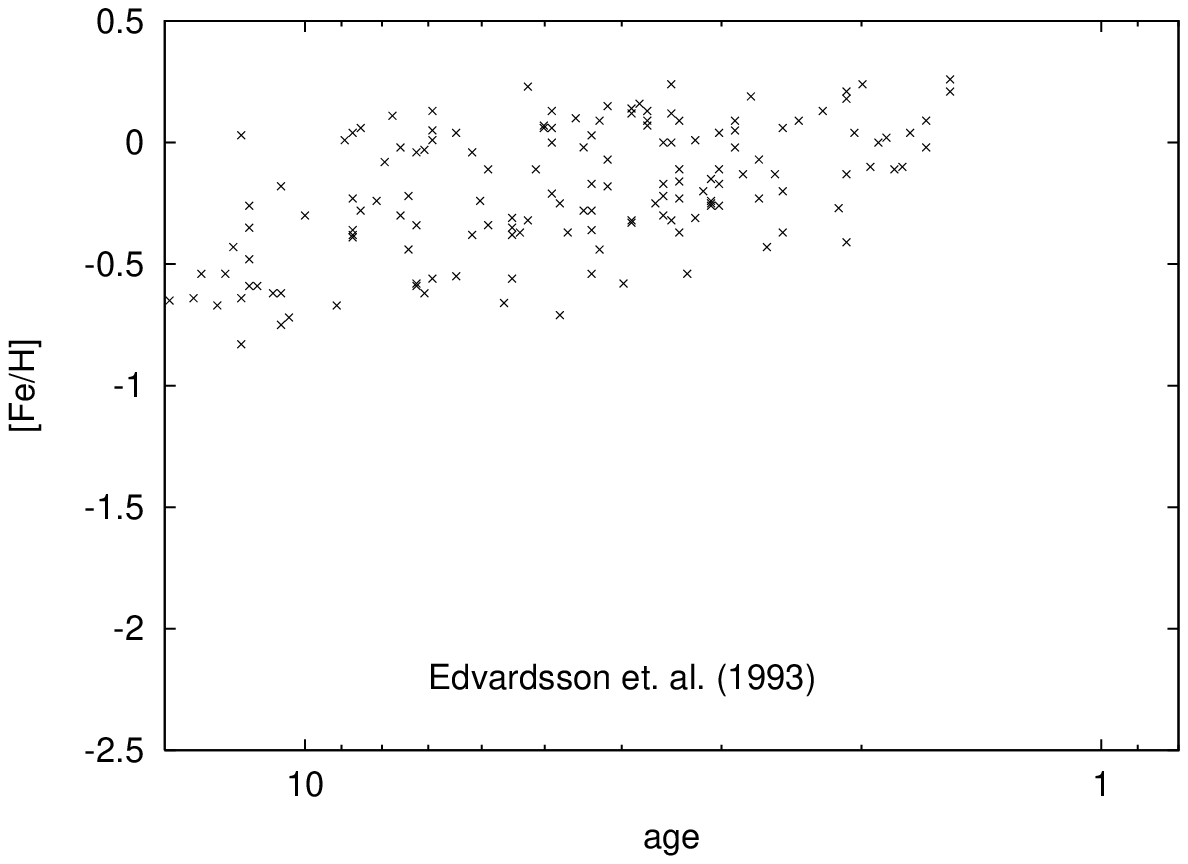}}
\caption{The photometric AMR for thin disc stars by \citet{Edvardsson_et.al.1993}.}
\label{Edamr}
\end{figure}

For thin disc stars, general trends of the AMR are apparently inconsistent 
with those previously derived by \citet{Twarog1980}, \citet{Edvardsson_et.al.1993}, and
 \citet{Ng_Bertelli1998}. In our data, the mean metallicity is almost constant from 
14 Gyr to 1 Gyr and the scatter along the AMR decrease into younger stars while 
the mean metallicity increases gradually from [Fe/H]$=-0.8$ at 16 Gyr to 
[Fe/H]$=-0.1$ at the present in the previous researches.
Since the Edvardsson et. al.'s AMR includes thick disc and halo stars in these previous researches, however, 
the direct comparison is not appropriate.
We carefully examined, therefore, both orbits and the AMR of \citet{Edvardsson_et.al.1993}. 
Figure \ref{Edamr} shows the AMR by \citet{Edvardsson_et.al.1993} for thin disc stars 
according to the criteria given in Sect. 2.2.2. 
We see that a similar feature is already seen in \citet{Edvardsson_et.al.1993} if we consider the stars younger than 10 Gyrs. For the stars older than 10 Gyrs, our AMR includes more metal rich stars than that of \citet{Edvardsson_et.al.1993}, and yet it would be difficult to discuss this difference in detail because the age determination of these old stars contains larger error than younger stars and the number of the sample stars of \citet{Edvardsson_et.al.1993} is not enough adequate. 
There are several features that were not clearly seen in Edvardsson et al.'s AMR: a) the upper 
envelope of the AMR is remarkably flat at a level similar to the solar
metallicity ([Fe/H]$\simeq 0.0 - 0.1$) for all the time from 14 Gyr
to 1 Gyr; and b)
although the AMR tends to converge to the point [Fe/H]$\simeq +0.3$ at
1.6 Gyr in the \citet{Edvardsson_et.al.1993}'s AMR, we found that 
a fairly large scatter in [Fe/H] still exists even at 1 Gyr.

%For stars older than 10 Gyr with $[Fe/H]\le -1.0$, on the other hand, 
%the scatter along the AMR is fairly large, which implies
%that chemical enrichment proceeded rather stochastically in the
%early stage of Galaxy evolution (e.g. \citet{Ikuta_Arimoto1999} ).

\subsection{The abundance pattern of thin disc} 

\begin{figure}
\resizebox{\hsize}{!}{\includegraphics{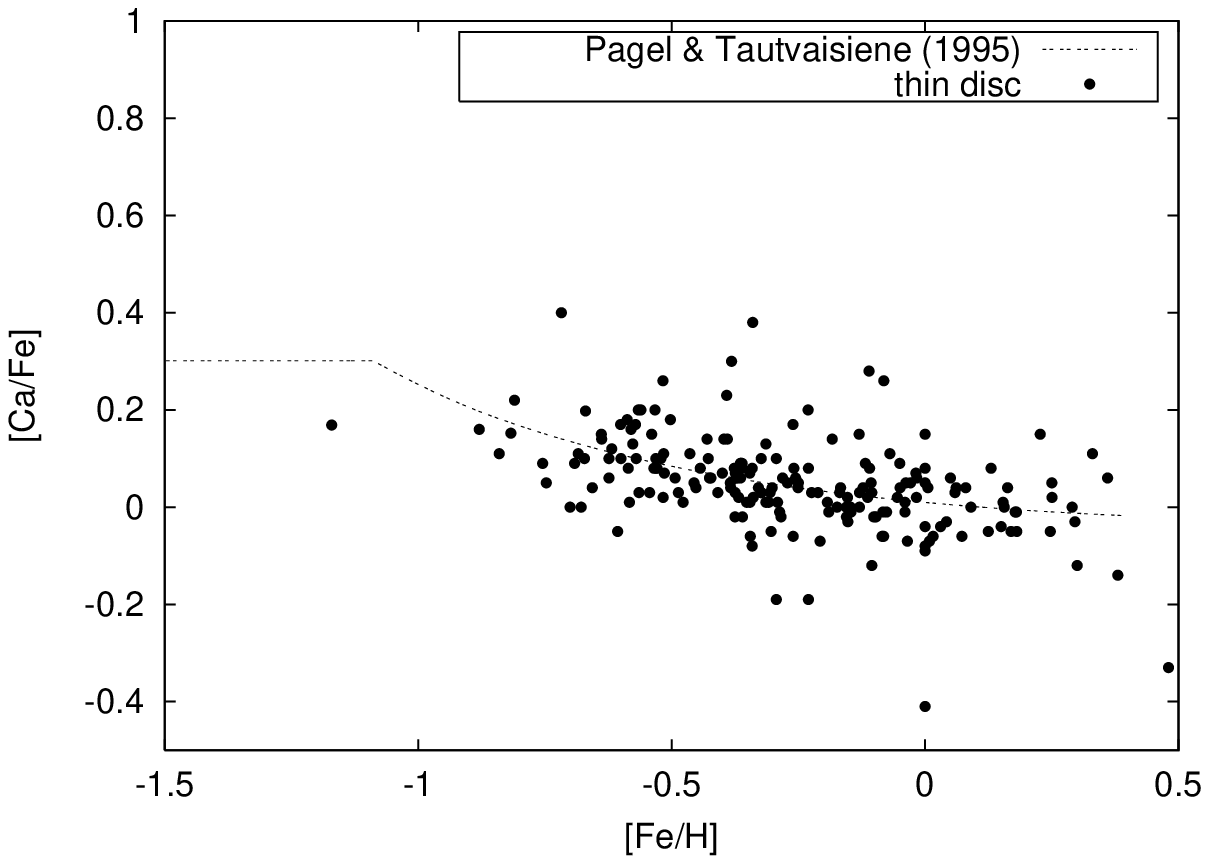}}
\caption{The [Ca/Fe] - [Fe/H] diagram for thin disc stars.}
\label{thin_CaFe_FeH}
\end{figure}

Figure \ref{thin_CaFe_FeH} shows the [Ca/Fe] - [Fe/H] relation for thin disc stars. 
The line represents the GCE model by \citet{Pagel_Tautvaisiene1995}. 
This model predicts that 
[Ca/Fe] $\simeq 0.3 $ for [Fe/H] $< -1.2$ (only SNII contribute for metal production) 
and that [Ca/Fe] 
decrease as [Fe/H] increase because of iron production from SNIa.  
The model shows good agreement with the averaged value of our data. 
Namely, our data shows the correlation that the more iron-rich stars show smaller [Ca/Fe].
The dispersion of [Ca/Fe] along the model line is, however, larger than that expected from 
uncertainty of observation. The scatters among the age, [Fe/H], and [Ca/Fe] are discussed in Sect. 5.5.

\section{The Thick Disc}
\subsection{The thick disc AMR}

\begin{figure*}
\sidecaption
\includegraphics[width=12cm]{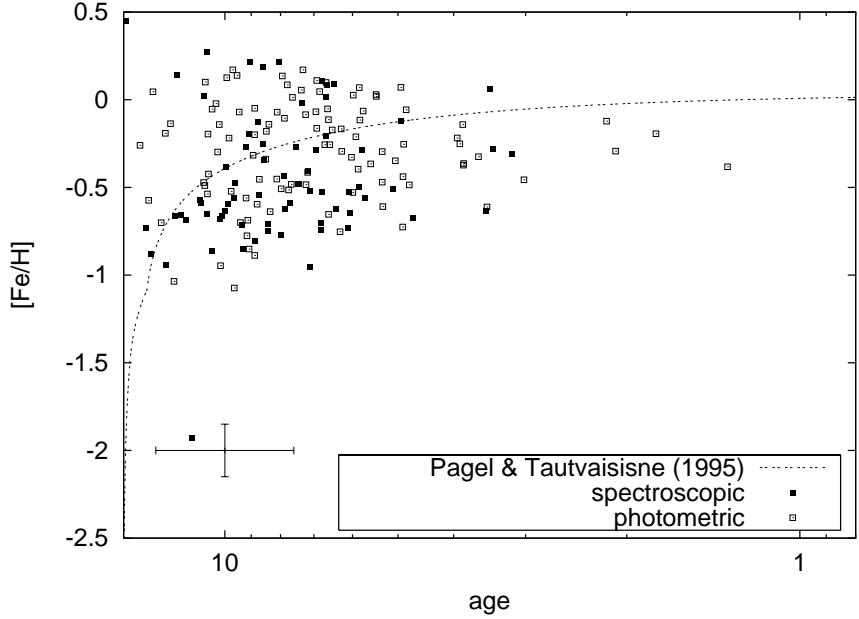}
\caption{The spectroscopic and photometric AMR for thick disc stars.}
\label{thick_AMR_spec}
\end{figure*}

Figure \ref{thick_AMR_spec} presents the spectroscopic AMR and the photometric AMR for thick disc. We realise three features, i.e.,

1) Bulk of thick disc stars are older than 5 Gyr. The average age of thick disc stars is 8.2 Gyr. 
The AMR suggests that the star formation in thick disc seized almost 8.2 Gyr ago.

2) The mean metallicity of thick disc stars is $\langle$ [Fe/H] $\rangle \sim -0.5 $ and the spread in [Fe/H] ranges from [Fe/H] = $-1.0$ to solar. This feature is consistent with previous researches \citep{Gilmore_Wyse1985,Carney_et.al.1989,Layden1995a,Layden1995b}

3) The scatter along the AMR is larger than that for thin disc.

\subsection{The abundance pattern of thick disc}

\begin{figure}
\resizebox{\hsize}{!}{\includegraphics{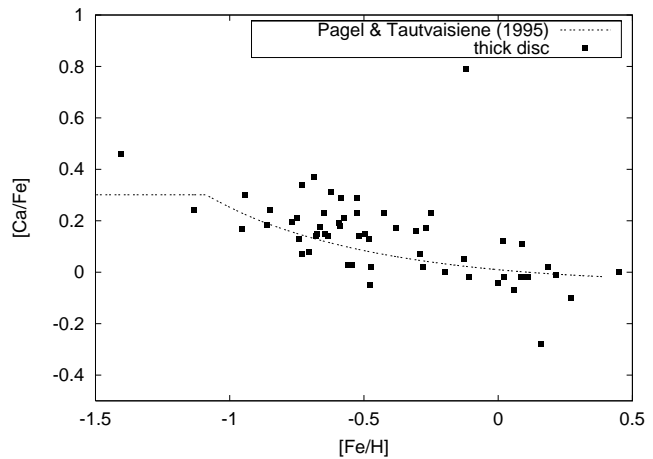}}
\caption{The [Ca/Fe] - [Fe/H] diagram for thick disc stars.}
\label{thick_FeH_CaFe}
\end{figure}

Figure \ref{thick_FeH_CaFe} show the [Ca/Fe] - [Fe/H] diagram for thick disc stars.
The dotted line in Fig.\ref{thick_FeH_CaFe} represents the best fit model for the chemical evolution for outer thin disc by \citet{Pagel_Tautvaisiene1995}. 
Clearly, Fig.\ref{thick_FeH_CaFe} shows that the thick disc stars are more $\alpha$-enhanced than the thin disc stars at the same [Fe/H] value.

This trend has been already reported by \citet{Gratton_et.al.2000} and \citet{Prochaska_et.al.2000}.
\citet{Prochaska_et.al.2000} suggested that $\alpha$-enhancement in the thick disc
indicates that the thin disc stars formed from gas more polluted by SNIa, which suggests 
that the thin disc formation significantly delayed ($>$ 1 Gyr) from the thick disc formation. 
We could not find, however, any clear evidence for delayed thick disc formation from the AMRs of thin
 disc and thick disc stars. 
Interestingly, bulge stars of the Galaxies show the similar trend of $\alpha$-enhancement and are believed to be experienced the rapid star formation history \citep{McWilliam_Rich1994,Rich_McWilliam2000}. 
The $\alpha$-enhancement of thick disc stars, therefore, may be also understood as the result of the rapid star formation.

\setcounter{table}{3}
\section{Discussion}
\subsection{Uncertainty in the AMR}
\subsubsection{Inhomogeneity of Spectroscopic [Fe/H] data}

Unfortunately, our AMR contains large errors in metallicity;
typically $\epsilon_{\rm [Fe/H]}=0.15$ dex due to inhomogeneous data 
taken from different authors, while $\epsilon_{\rm [Fe/H]}=0.1$ 
dex in \citet{Edvardsson_et.al.1993} due to their homogeneous [Fe/H]
data reduced by the same analysis method. 
Even if stellar spectra were taken with
high S/N and high resolution, [Fe/H] estimates by different authors
could result in large scatter in [Fe/H].

Therefore, without examining carefully 
the details of individual analyses to understand the differences,
it would be dangerous to argue too much details of the value of metallicity.
We also have to keep in our mind that [Fe/H] determinations
are affected by the adopted effective temperatures, gravities and
microturbulent velocities, and that a stellar metal abundance
can be in error, even if the observations are of excellent
quality \citep{CayrelCatalogue1997}. 

However, recent observations are in good agreement for different observers.
We have studied the 
most observed 30 stars in the [Fe/H] catalogue to find the 
average dispersion in metallicity is 0.13 dex (See. Table \ref{var_table}). Our sample includes 
extremely metal-deficient stars, in which the dispersion in metallicity is apt to show the larger values. The dispersion in the metallicity for stars with [Fe/H] $>$ -2.5 is even smaller. 
Considering that the our sample includes a very small number of extremely metal-deficient stars, we conclude that the dispersion in the metallicity can be estimated to be 0.11 dex.

\begin{table*}
\caption{Differences of [Fe/H] values with different observations.}
\small
\begin{tabular}{lrrrrr}\hline\hline
Name & Number of obs. & [Fe/H] $_{\rm average}$ & [Fe/H]$_{\rm max}$ & [Fe/H]$_{\rm min}$ & $\epsilon_{\rm [Fe/H]}$ \\\hline
CD $-$38 245        &	8&	-3.95&	-4.42&	-3.00&	0.39\\
HD 122563           &	16&	-2.69&	-2.93&	-2.45&	0.10\\
BD +03 740         &	11&	-2.66&	-2.98&	-2.3&	0.22\\
\\
HD 140283           &	26&	-2.49&	-2.75&	-2.21&	0.14\\
BD +2 3375         &	10&	-2.28&	-2.65&	-1.95&	0.19\\
HD 216143           &	9&	-2.21&	-2.26&	-2.10&	0.05\\
HD 128279           &	9&	-2.19&	-2.50&	-1.97&	0.16\\
HD 165195           &	11&	-2.18&	-2.26&	-1.92&	0.09\\
HD 84937           &	12&	-2.14&	-2.43&	-1.86&	0.17\\
BD +37 1458         &	9&	-2.02&	-2.33&	-1.79&	0.19\\
\\
HD 19445           &	19&	-2.00&	-2.31&	-1.76&	0.14\\
%HD 074000           &	9&	-1.96&	-2.23&	-1.52&	0.19\\
%HD 204543           &	9&	-1.82&	-1.96&	-1.64&	0.0939\\
%HD 160617           &	8&	-1.79&	-2.04&	-1.48&	0.182\\
%HD 025329           &	8&	-1.78&	-1.94&	-1.67&	0.0918\\
HD 187111           &	13&	-1.78&	-2.35&	-1.54&	0.19\\
HD 26297           &	10&	-1.77&	-1.87&	-1.68&	0.06\\
HD 122956           &	13&	-1.75&	-1.96&	-1.53&	0.10\\
HD 64090           &	12&	-1.68&	-1.94&	-1.49&	0.16\\
%HD 044007           &	9&	-1.62&	-1.71&	-1.25&	0.147\\
HD 211998           &	10&	-1.53&	-1.68&	-1.25&	0.14\\
\\
%HD 045282           &	8&	-1.47&	-1.58&	-1.29&	0.0923\\
HD 83212           &	11&	-1.46&	-1.57&	-1.37&	0.06\\
%HD 175305           &	9&	-1.45&	-1.55&	-1.3&	0.0704\\
%HD 345957           &	8&	-1.45&	-1.68&	-1.3&	0.123\\
HD 94028           &	11&	-1.44&	-1.66&	-1.31&	0.12\\
HD 103095           &	15&	-1.34&	-1.59&	-1.17&	0.11\\
%HD 111721           &	8&	-1.33&	-1.68&	-0.98&	0.198\\
%HD 166161           &	9&	-1.28&	-1.54&	-1.11&	0.121\\
%HD 003567           &	9&	-1.25&	-1.48&	-1.05&	0.121\\
HD 194598           &	17&	-1.14&	-1.37&	-0.99&	0.11\\
%HD 193901           &	9&	-1.08&	-1.22&	-0.91&	0.113\\
HD 201891           &	15&	-1.01&	-1.12&	-0.87&	0.06\\
\\
%BD +29 0366         &	8&	-0.972&	-1.14&	-0.78&	0.118\\
%HD 064606           &	8&	-0.94&	-1.01&	-0.83&	0.061\\
%HD 025704           &	11&	-0.935&	-1.13&	-0.64&	0.146\\
HD 76932           &	14&	-0.92&	-1.05&	-0.76&	0.09\\
HD 63077           &	15&	-0.89&	-1.16&	-0.53&	0.16\\
HD 022879           &	12&	-0.86&	-0.99&	-0.76&	0.05\\
%HD 134169           &	10&	-0.819&	-0.98&	-0.68&	0.0967\\
HD 59984           &	14&	-0.82&	-1.60&	-0.52&	0.24\\
%HD 106516           &	11&	-0.765&	-0.98&	-0.61&	0.0991\\
%HD 203608           &	11&	-0.719&	-0.92&	-0.57&	0.112\\
%HD 114762           &	11&	-0.715&	-0.87&	-0.56&	0.0881\\
%HD 184499           &	9&	-0.628&	-0.8&	-0.51&	0.109\\
%HD 207978           &	8&	-0.578&	-0.66&	-0.45&	0.068\\
%HD 010700           &	8&	-0.541&	-0.66&	-0.38&	0.0837\\
HD 124897           &	14&	-0.54&	-0.69&	-0.37&	0.10\\
%HD 216385           &	10&	-0.325&	-0.53&	-0.2&	0.108\\
%HD 000400           &	8&	-0.28&	-0.41&	-0.15&	0.0766\\
%HD 222368           &	10&	-0.232&	-0.45&	-0.1&	0.0964\\
%HD 020807           &	8&	-0.21&	-0.42&	0.1&	0.154\\
%HD 173667           &	9&	-0.0967&	-0.4&	0.03&	0.115\\
%HD 039587           &	10&	-0.025&	-0.18&	0.11&	0.0795\\
HD 61421           &	13&	-0.02&	-0.18&	0.05&	0.05\\
%HD 095128           &	8&	-0.0125&	-0.12&	0.01&	0.0409\\
\\
HD 9826           &	8&	0.09&	-0.03&	0.17&	0.05\\
HD 217014           &	8&	0.15&	0.05&	0.21&	0.07\\
HD 128620           &	10&	0.19&	0.10&	0.25&	0.04\\
 \hline
\end{tabular}

\normalsize
\label{var_table}
\end{table*}

\subsubsection{Uncertainty in the age}
\begin{figure*}
\centering
\includegraphics[width=17cm,height=6cm]{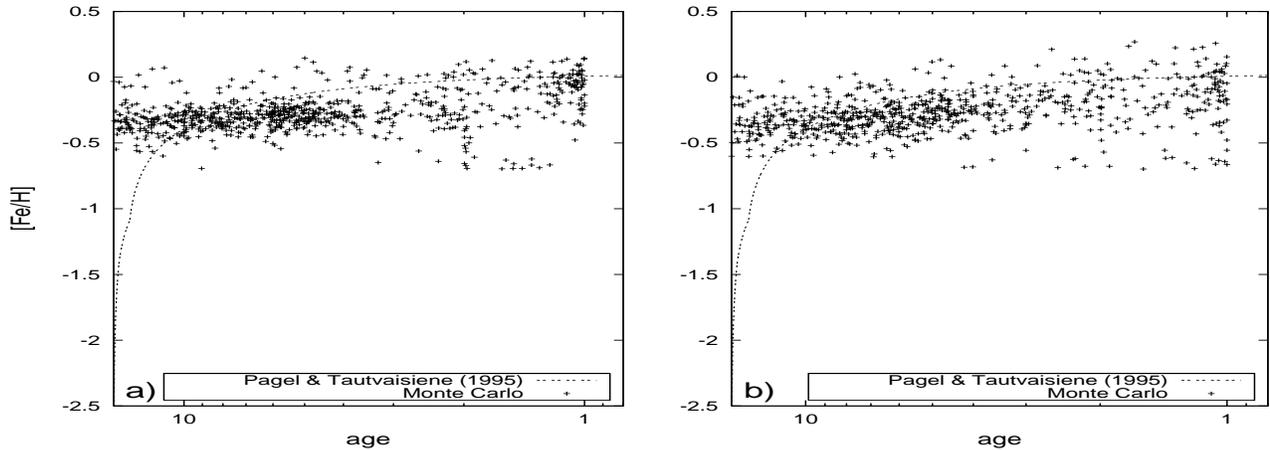}
\caption{The simulated AMR from one zone infall model by \citet{Pagel_Tautvaisiene1995}. \textbf{a)} The errors in $Z$ are set to be 0.07. \textbf{b)} these are set to be 0.13. The errors in $M_{V_0}$ and $(B-V)_0$ are 0.2 mag and 0.02 mag in both figures.}
\label{mc}
\end{figure*}

Uncertainties in $M_{V_0},(B-V)_0$ and $Z$ cause errors for resulting ages. Errors in ${M_V}_0$ and $(B-V)_0$ are typically 0.2 mag ($\simeq$ 10 \% in parallax) and 0.02 mag in our data. Combined with $Z$ uncertainties, we can determined the expected error in the resulting age. 
The errors in ages also depends on the initial mass of the stars through their position of colour-magnitude diagram and the nature of the isochrones hereon.  
We conducted, therefore, Monte Carlo simulations assuming the AMR and the star formation history in the one zone infall model of \citet{Pagel_Tautvaisiene1995}. 
We took the set of stars (the simple stellar population) from 15 Gyr old to the present, with given metallicity from AMR and star formation history of \citet{Pagel_Tautvaisiene1995}. 
The locations on the colour-magnitude diagram are derived from \citet{Yi_et.al.2001}, adding Gaussian errors of $M_{V_0},(B-V)_0$ and $Z$.
Then we obtained the age by using the same method in Sect. 2.2. 
Figures \ref{mc} shows the resulting AMR with errors of 0.07 in $Z$ (left) and 0.13 in $Z$ (right). Notably, the expected spread in age -- due only to errors -- shown by both panels of Fig. \ref{mc} is smaller than the observed dispersion of points in Fig. \ref{thin_AMR_spec} for stars older than 3 Gyr.

\subsection{Abundance Gradient}

%Radial and vertical abundance gradient is a key to the understanding of the Galaxy. 
%The first research of abundance gradient using stars is back to that of \citet{Mayor1976}.
In order to discuss the radial abundance gradient of the Galaxy, it is 
not appropriate to use the present position of the stars as mean position of the stellar orbit.
We used, instead, $R\equiv (R_p + R_a)/2$, for the distance from the Galactic centre \citep{Edvardsson_et.al.1993}. 
We also employed $z_{\rm max}$ for the characteristic distance from the Galactic plane.

\subsubsection{Thin disc abundance gradient}

\begin{figure}
\resizebox{\hsize}{!}{\includegraphics{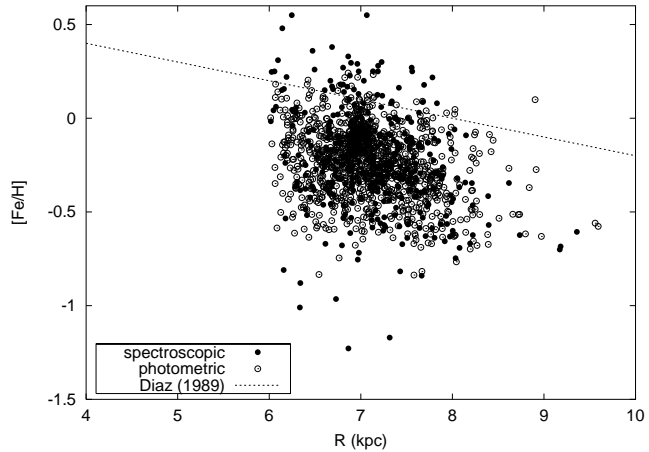}}
\caption{The [Fe/H]-$R$ relation for thin disc stars. Dashed line represents observations of \ion{H}{ii} regions \citep{Diaz1989}.}
\label{RFeH_thin}
\end{figure}

\begin{figure}
\resizebox{\hsize}{!}{\includegraphics{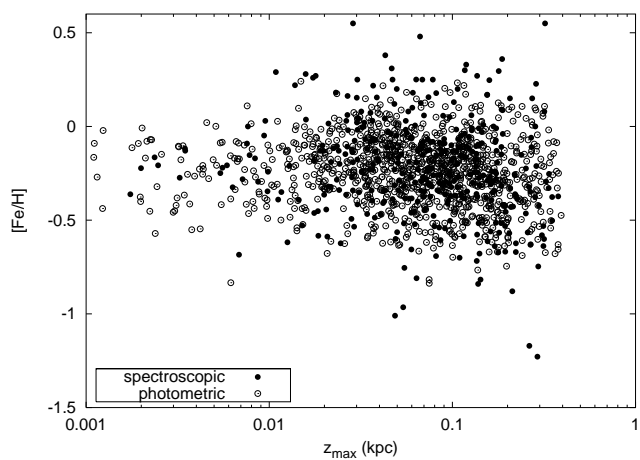}}
\caption{The [Fe/H]-$z_{\rm max}$ relation for thin disc stars.}
\label{ZFeH_thin}
\end{figure}

Figure \ref{RFeH_thin} presents the [Fe/H]-$R$ relation for thin disc stars. Filled symbols represent spectroscopic data while empty symbos represent shows photometric data. Both spectroscopic and photometric data shows no strong correlations between $R$ and $[Fe/H]$. The line are the observations derived by \citet{Diaz1989}. Figure \ref{ZFeH_thin} shows the [Fe/H] -$z_{\rm max}$ relation for thin disc stars. These data also show no strong correlations. 


\subsubsection{Thick disc abundance gradient}

\begin{figure}
\resizebox{\hsize}{!}{\includegraphics{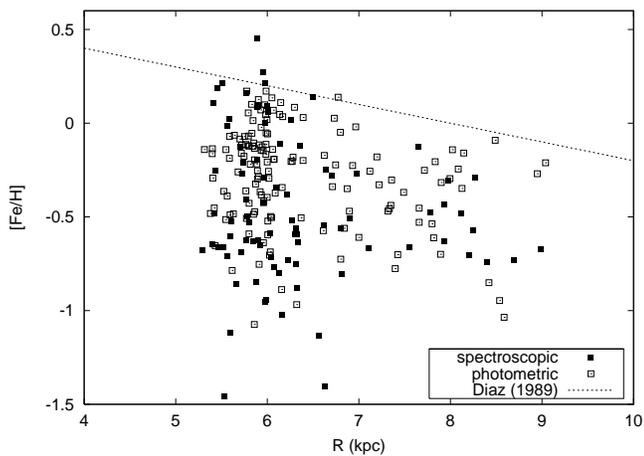}}
\caption{The [Fe/H]-$R$ relation for thick disc stars. Dashed line represents observations of \ion{H}{ii} regions \citep{Diaz1989}.}
\label{RFeH_thick}
\end{figure}

\begin{figure}
\resizebox{\hsize}{!}{\includegraphics{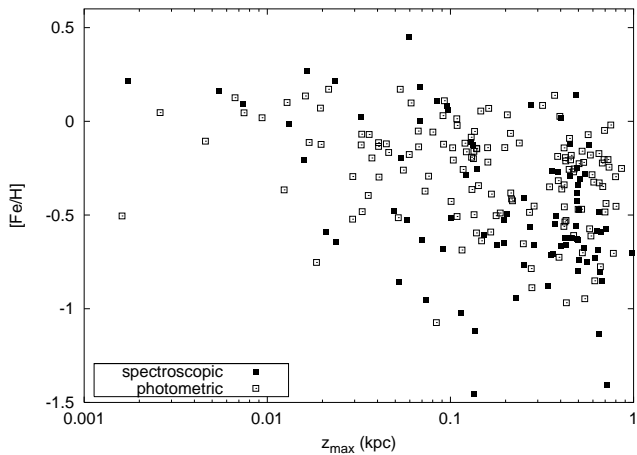}}
\caption{The [Fe/H]-$z_{\rm max}$ relation for thick disc stars.}
\label{ZFeH_thick}
\end{figure}

Figure \ref{RFeH_thick} represents the [Fe/H]-$R$ relation for thick disc stars. Filled symbols represent spectroscopic data, while empty ones represent photometric data. Both spectroscopic and photometric data show little correlation between $R$ and [Fe/H].  Figure \ref{ZFeH_thick} shows the [Fe/H] - $z_{\rm max}$ relation for thick disc stars. Unlike thin disc stars, thick disc stars show the feature that the stars more distant from the Galactic plate tend to be more metal poor. Namely, a vertical abundance gradient is seen in thick disc. Comparing the thin disc stars and thick disc stars with 100 pc $< z_{\rm max} <$ 600 pc, the thick disc stars are clearly more metal-deficient than the thin disc stars. This phenomenon is naturally understood if we surmise that these two kinds of stellar groups have different origin: the thick disc stars located at the distance $z_{\rm max} >$ 100 pc formed intrinsically in the region distant from the Galactic plane while the thin disc stars formed near the Galactic plane and drifted or heated-up vertically to 100 pc away from their original birth place.    

\subsection{The G-dwarf problem}

\begin{figure}
\resizebox{\hsize}{!}{\includegraphics{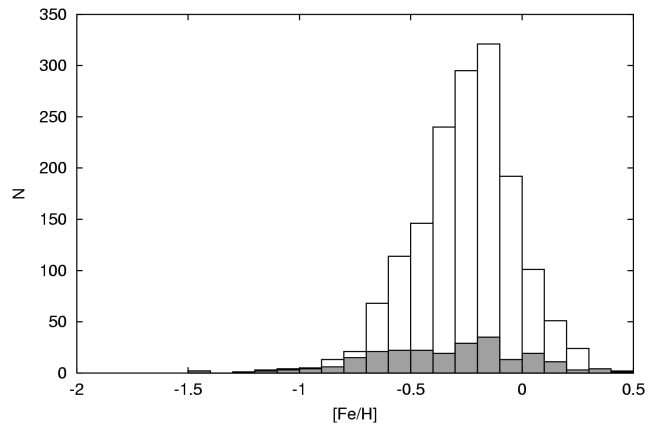}}
\caption{Abundance distribution function for thin disc (open boxes) and thick disc (gray boxes).}
\label{mdf}
\end{figure}
Figure \ref{mdf} shows the metallicity distribution function for thin disc stars and thick disc stars. Although it is impossible to discuss the property of the abundance distribution because our sample lacks completeness, it is clear that both thin disc and thick disc are deficient in metal-poor stars.
Notably, the G-dwarf problem do exist for thick disc stars. Since the metallicity distribution function of metal deficient halo stars can be explained by the simple model \citep{Laird_et.al.1988}, thick disc presents a striking contrast to halo, instead abundance distribution function for thick disc stars shows rather close feature to that of bulge \citep{McWilliam_Rich1994}.

\subsection{Formation process for thick disc}

We here discuss how thick disc formed in the Galaxy. 
In summary, thick disc shows three important features, i.e. 
1) thick disc stars are older than 5 Gyr, 
2) thick disc stars show vertical abundance gradient, and 
3) thick disc stars have different abundance patterns compared with thin disc counterparts. 
To date, thick disc formation scenarios, such as monolithic collapse \citep{Larson1976,Jones_Wyse1983}, dynamical heating \citep{Noguchi1998}, major merger, and accretion of dwarf galaxies \citep{Quinn_Goodman1986} have been proposed.

Major merger should invoke a strong peak  in age-distribution. It is also widely believed that 
merger will work as reducing vertical abundance gradient. Our research does not, therefore, support 
the thick disc formation with major merger.   
If thick disc is formed from heating of thin disc, the star formation history of thick disc should be similar to that of thin disc and the similar star formation history essentially produces the similar [Ca/Fe]-[Fe/H] relation. Heating mechanism is, therefore, unlikely to explain the thick disc formation.
The other two theories both account for our result.

\subsection{Scatter in the AMR for thin disc}	

In this section, we will try to find the way to explain the scatter along the AMR by simple extension of the one zone model derived by \citet{Pagel_Tautvaisiene1995}.

\subsubsection{Metallicity-Orbit relation}
\begin{figure*}
\centering
\includegraphics[width=17cm,height=14cm]{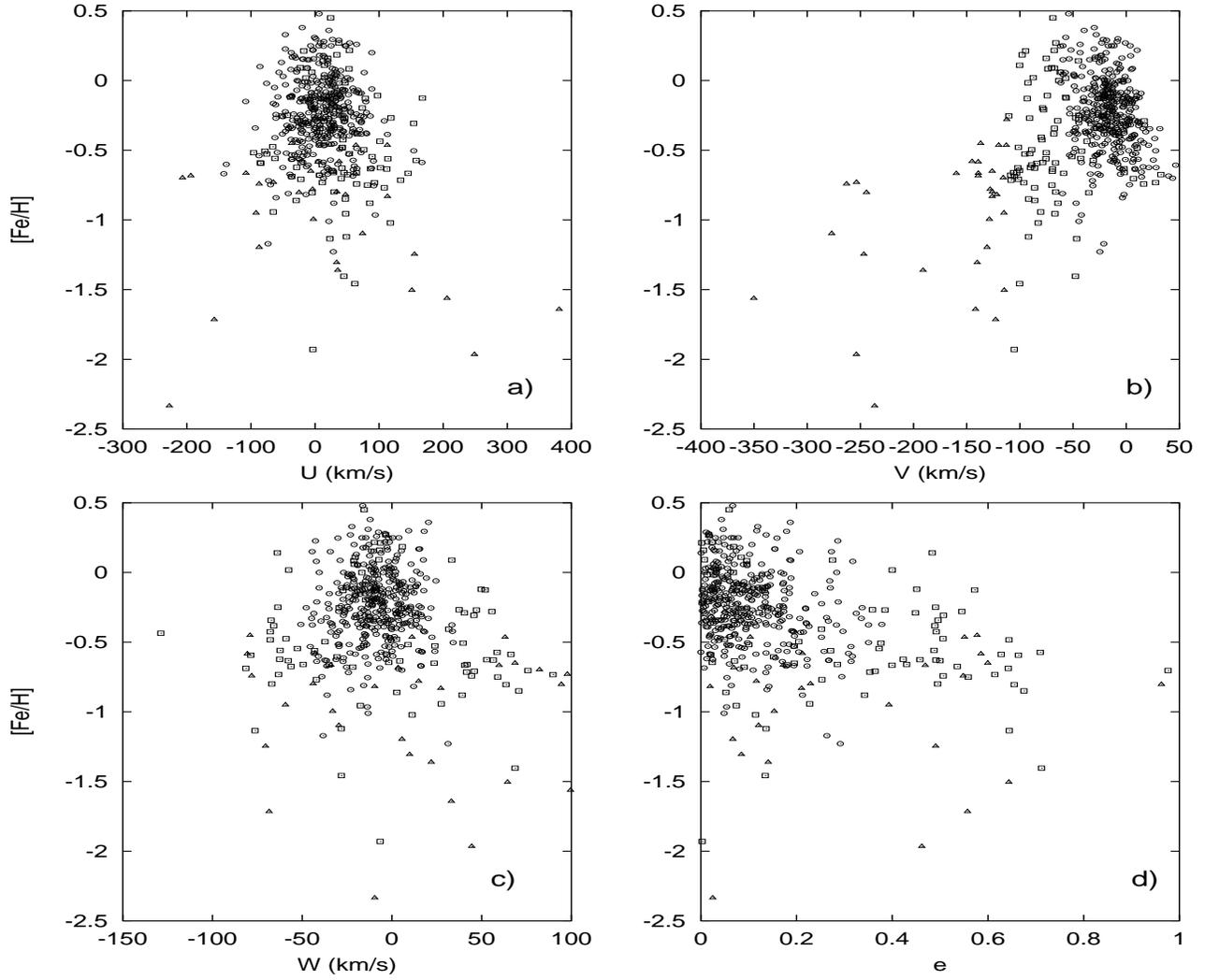}
\caption{\textbf{a)} $U$ - [Fe/H]  relation. \textbf{b)} $V$-[Fe/H]. \textbf{c)} $W$-[Fe/H]. \textbf{d)} $e$-[Fe/H]. Circles, squares, and triangles represent thin disc, thick disc, and halo stars, respectively.}
\label{orbit}
\end{figure*}

The solar neighbourhood stars consist of a mixture of stars born at different place with different orbit. This fact may account for the scatter along the thin disc AMR \citep{Edvardsson_et.al.1993,Pagel_Tautvaisiene1995,Prantzos_Boissier2000}. 
Already seen in Sec. 5.2, the abundance gradient is not significant for thin disc stars. Another relation between orbital elements and metallicity relation may exist for thin disc stars. For example, if we find metallicity-orbit relation that stars with larger eccentricity tend to be more metal deficient, the scatter along the AMR can be attributed to such kind of relation. Therefore, we examined the $U$, $V$, $W$, $e$ - [Fe/H] relation (Fig. \ref{orbit}). All of these relations, however, show more a scattered feature than a strong correlation. Notably, a large scatter in [Fe/H] is seen even if the stars with the same orbital parameters are selected. The scatter along the AMR cannot, therefore be attributed to the relation of this kind.

\subsubsection{Inhomogeneous effective yield and star formation}
The effective yield can be different from place to place although the true yield is homogeneous, because the supernova ejecta are inhomogeneous due to axisymmetric explosions of rotating massive stars \citep{Maeda_et.al.2001}. 
Even if the explosions are isotropic, the ejecta may stochastically contaminate the surroundings, 
unless the  ISM is distributed uniformly. 
The scatter can be explained even if the star formation history in the ``solar torus'' is completely homogeneous. 
It is also possible that the local inhomogeneity of star formation rate produces the scatter. 
To examine these scenarios, we prepared 7 models by modifying the parameters in the analytical model of \citet{Pagel_Tautvaisiene1995}.
Table \ref{models} shows the adopted parameters for each model. Model A is an original infall model taken from \citet{Pagel_Tautvaisiene1995}. Model B and C assumed twice and half yield SNII ejecta, respectively. SNIa yield and star formation timescale is changed in model D, E and model F, G in the same manner.

\begin{table}
\caption{The parameters for 7 GCE models, originally introduced by \citet{Pagel_Tautvaisiene1995}. $1/\omega$ represent the star formation timescale.}
\begin{tabular}{llrrrrr}\hline
model & $\omega$ & $y_{\rm Fe,SNIa} $& $y_{\rm Fe,SNII }$ & $y_{\rm Ca,SNIa} $ &  $y_{\rm Ca
,SNII} $\\ \hline
A& 0.3 & 0.42 & 0.28 & 0.18 & 0.56\\
B& 0.3 & 0.42 & 0.56 & 0.18 & 1.12 \\
C& 0.3 & 0.42 & 0.14 & 0.18  & 0.28 \\
D& 0.3 & 0.84 & 0.28 & 0.36 & 0.56 \\
E& 0.3 & 0.21 & 0.28 & 0.09 & 0.56 \\
F& 0.15 &  0.42 & 0.28 & 0.18 & 0.56 \\
G& 0.6& 0.42 & 0.28 & 0.18 & 0.56\\
\hline
\end{tabular}
\label{models}
\end{table}

\begin{figure*}
\centering
\includegraphics[width=17cm]{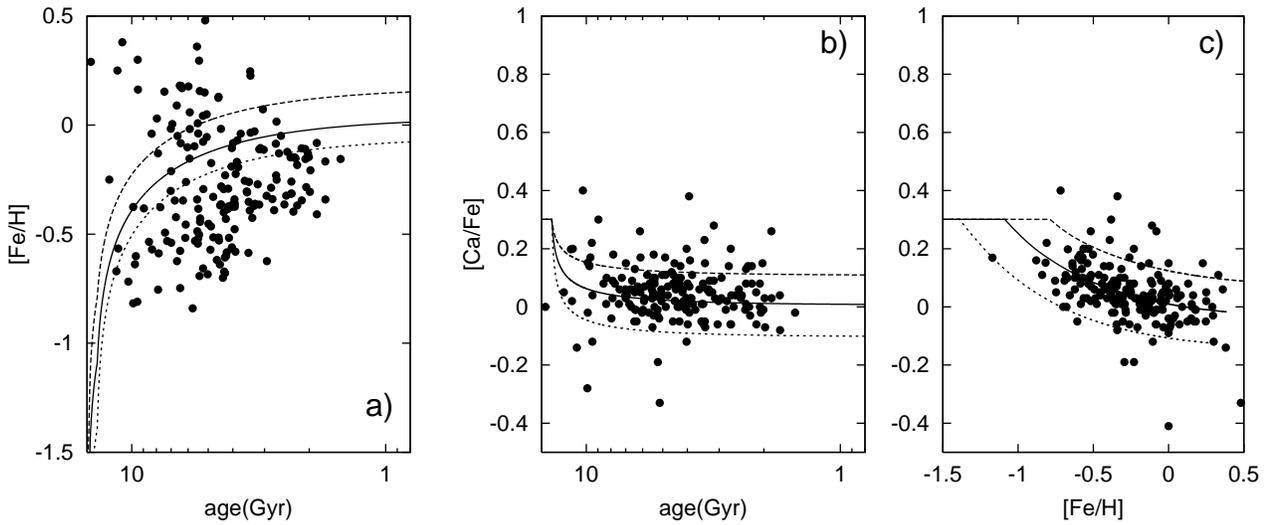}
\caption{Comparison of \textbf{a)}the AMR, \textbf{b)}age-[Ca/Fe] relation, and \textbf{c)} [Ca/Fe]-[Fe/H] diagram with models A (solid line), B (long dashed line), and C (short dashed line).}
\label{yII}
\end{figure*}

The scatter may arise if relative contributions of SNIa and SNII differ from place to place.
The various panels of Fig. \ref{yII} show the AMR, age-[Ca/Fe] relation, and the [Ca/Fe]-[Fe/H] diagram with models A, B, and C. 
Clearly the difference in the effective yield of SNII produces smaller scatter along the AMR than the observation, while the scatter in the [Ca/Fe]-[Fe/H] diagram (right) is too large to explain the observation. 

\begin{figure*}
\centering
\includegraphics[width=17cm]{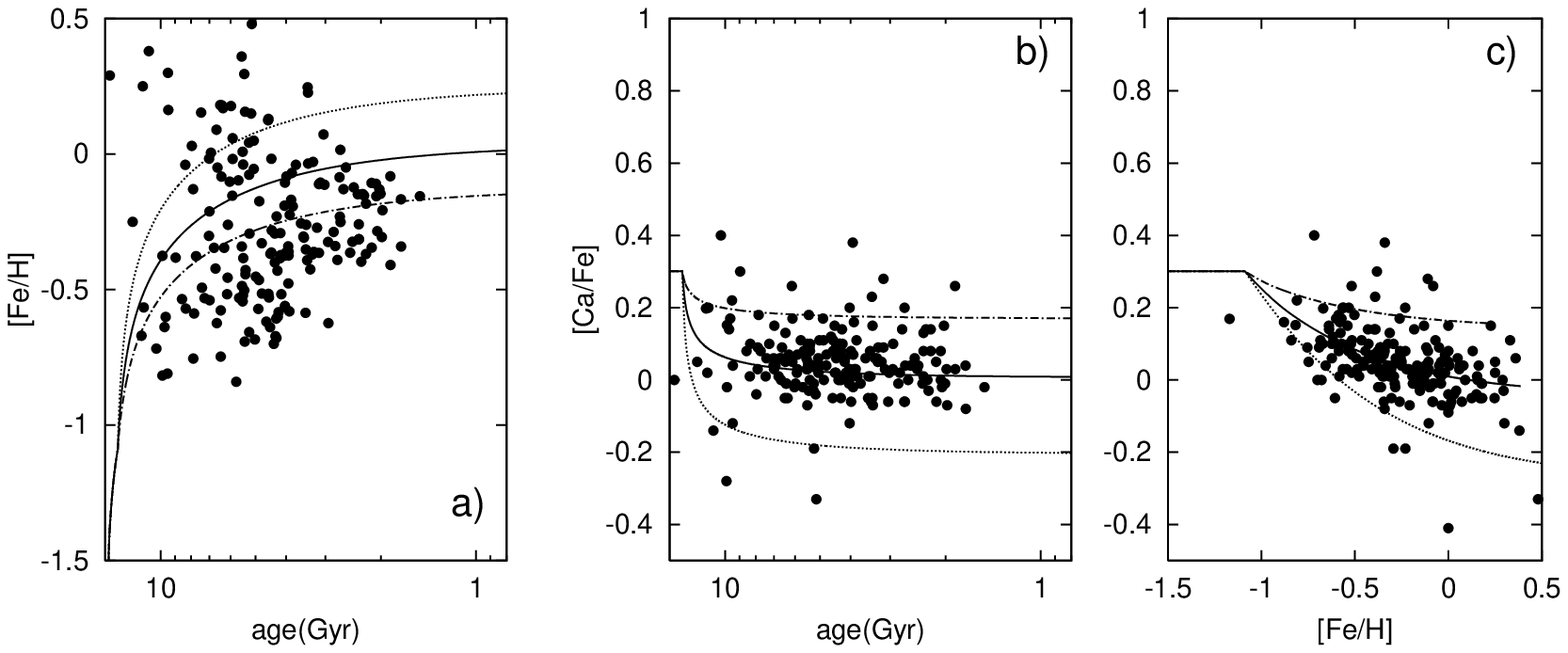}
\caption{Comparison of \textbf{a)} the AMR, \textbf{b)} age-[Ca/Fe] relation, and \textbf{c)} [Ca/Fe]-[Fe/H] diagram with models A (solid line), D (dotted line), and E (dash-dotted line).}
\label{yI}
\end{figure*}

\begin{figure*}
\centering
\includegraphics[width=17cm,height=6cm]{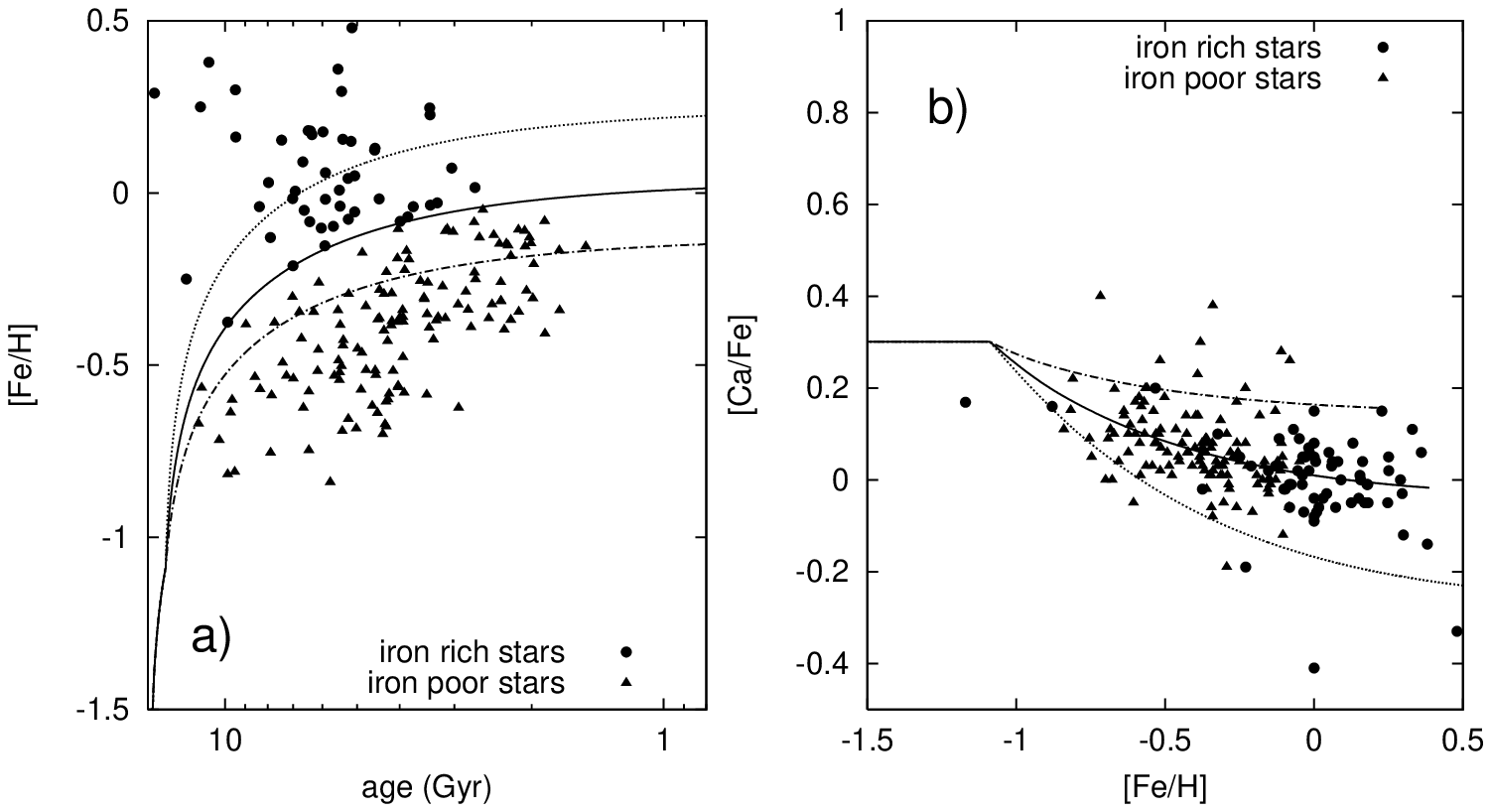}
\caption{Comparison of \textbf{a)} the AMR and  \textbf{b)} the [Ca/Fe] diagram in detail. Circles represent stars more iron rich than the model A, while triangles represent iron poor stars.}
\label{yI+}
\end{figure*}

Next, we examined the inhomogeneous effective yield of SNIa.
Figures \ref{yI} show the models with different effective yields for SNIa. 
Models A,D, and E reproduce fairly well 
both the trend of thin disc stars' AMR and the scatter along it. 
However, dividing the [Fe/H] relation into two parts (Fig. \ref{yI+}), metal rich stars and metal poor stars, and examining the [Ca/Fe] -[Fe/H] diagram in detail, we found that this model cannot acount for the scatters. In Fig. \ref{yI+} the sample stars are divided into two parts. More iron rich stars than model A are designated in circles while metal poor stars are in triangles. Model D and E predict that iron rich stars (circles) will show smaller [Ca/Fe] than model A on Fig. \ref{yI+} and that the metal poor stars (triangles) will show larger [Ca/Fe] than model A. 
In our data, however, both iron rich stars and iron poor stars are distributed along model A and show no clear split of [Ca/Fe] on Fig. \ref{yI+}.  
Thus, models with different effective yield for SNIa cannot explain this feature on the [Ca/Fe] - [Fe/H] diagram.

\begin{figure*}
\centering
\includegraphics[width=17cm]{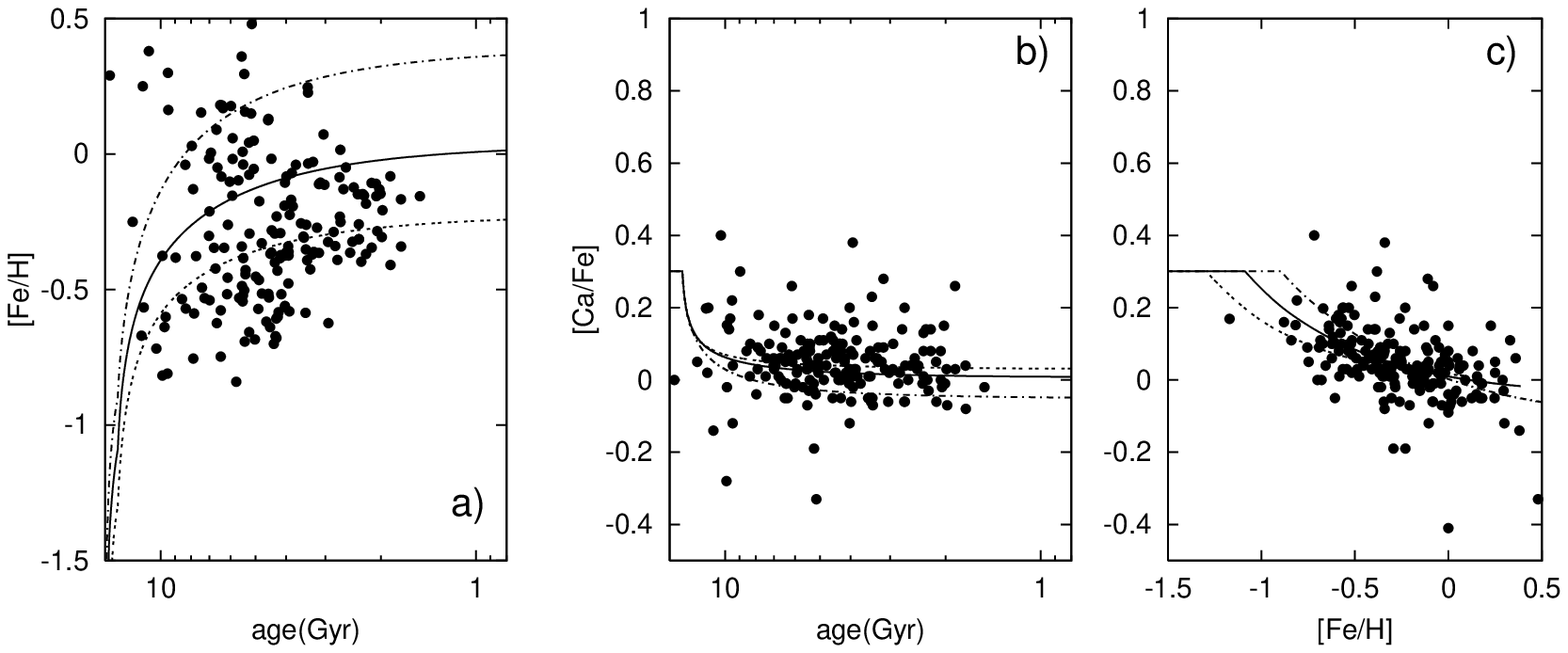}
\caption{Comparison of \textbf{a)} the AMR, \textbf{b)} age-[Ca/Fe] relation, and \textbf{c)} [Ca/Fe]-[Fe/H] diagram with models A (solid line), F (dash-dotted line), and G (short dashed line).}
\label{sfr}
\end{figure*}

Finally, the inhomogeneity of star formation rate is examined. 
Figures \ref{sfr} show the models with different star formation rate.
The scatter along the age-[Ca/Fe] relation is cleary larger than expected from the models.

Comparing these three hypotheses and our data, we conclude that such a simple modification of one-zone model cannot explain the scatter along the AMR, age-[Ca/Fe] relation, and [Ca/Fe]-[Fe/H] relation simultaneously.

\subsubsection{Planet migration}
\begin{figure}
\resizebox{\hsize}{!}{\includegraphics{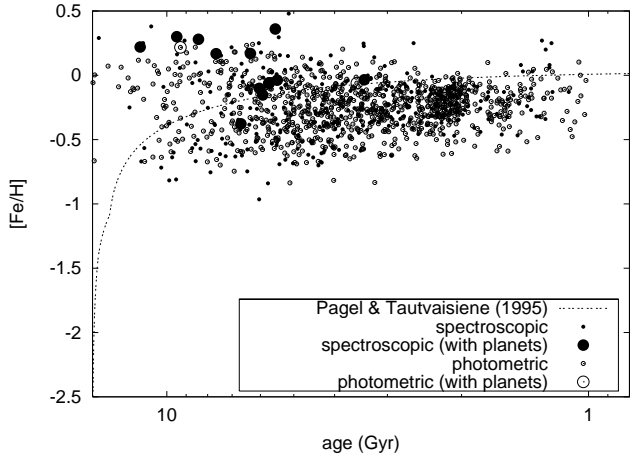}}
\caption{The AMR and planets hosting stars in our sample. Large symbols denote known planets host stars.}
\label{planets}
\end{figure}

There is a growing evidence that the star formation mechanism may have some influence on the metallicity of the star: in particular, \citet{Gonzalez1997} has recently suggested that stars with planets are systematucally more metal-rich than stars without planets, and that this is due to the presence of planets (``planet migration''). 
This effect can cause the scatter in the AMR. Our data include, however, only 14 discovered planet host stars \citep{Mayor_Queloz1995,Fischer_et.al.1999,Butler_et.al.2000,Henry_et.al.2000,Korzennik_et.al.2000,Marcy_et.al.2000,Mazeh_et.al.2000,Udry_et.al.2000,Vogt_et.al.2000,Butler_et.al.2001,Fischer_et.al.2001,Gonzalez_et.al.2001,Tinney_et.al.2001}, and we have scatter in the AMR when we consider only stars without planets. Fig. \ref{planets} shows the AMR with known planets host stars and other stars. As already noted by \citet{Gonzalez1997} the planets host stars tend to have larger metallicities than the general trend of the AMR. It should be noted that the Sun also shows this ``over-metallicity''. 
It seems that the idea is reasonable because the ``over-metallcity'' of 0.1 dex in [Fe/H] result in the underestimate of the age of $\sim$ 1 Gyr. If we consider that ages and [Fe/H] of planets host stars are underestimated and oversetimated, respectively, the original [Fe/H] and ages of planets stars shows better fit to the one-zone AMR model.
However, it is insufficient or even dangerous to study the general nature of the scatter of the AMR from such small number of stars (less than 1\% of the total samples stars) and we cannot conclude yet that the planet migration is the major cause for the scatter along the AMR.

\section{Summary \& Conclusion}
The AMR and orbital parameters are newly derived for 1658 solar neighbourhood stars to which accurate distances are measured by the {\it HIPPARCOS} satellite. 
The sample stars are divided into 1382 thin disc stars, 229 thick disc stars, and 47 halo stars according to their orbital parameters.
Notably, the thin disc AMR shows a considerable scatter along the one-zone GCE model. 
No clear relation between orbit and metallicity are found. Namely, the scatter along the AMR exists even if the stars with the same orbits are selected. 
We examined simple extension of one-zone GCE models which account for i) inhomogeneity in the effective yield caused by the spatially localised mixing of inter stellar medium; and ii) inhomogeneous star formation history. 
We found both extensions of one-zone GCE model cannot account for the scatter in age - [Fe/H] - [Ca/Fe] relation simultaneously. 
In our work, the scatter along the AMR for thin disc stars have been confirmed for far larger samples (1382 stars) , which is more than 5 times larger than that used in the previous work of \citet{Edvardsson_et.al.1993}. 
We concluded that this scatter, which should be accounted for by any Galaxy models, is one of the most important feature in the formation and evolution of the Galaxy. 
On the other hand, the AMR for thick disc stars shows that the star formation terminated 8 Gyr ago in thick disc. 
We reconfirmed the trend that the thick disc stars are more Ca-rich than the disc stars with the same [Fe/H], which has been already reported by \citet{Gratton_et.al.2000} and \citet{Prochaska_et.al.2000}.
Thick disc stars show a vertical abundance gradient. 
These three facts, the AMR, vertical gradient, and [Ca/Fe]-[Fe/H] relation, support monolithic collapse and/or accretion of satellite dwarf galaxy as thick disc formation scenario.

\acknowledgement{We are grateful to the anonymous referee whose suggestions on the early version greatly improved the paper. We thank to Dr. Chiaki Kobayashi for useful discussions. A. I. thanks the Japan Society for Promotion of Science (JSPS) Research Fellowships for Young Scientists.}

\setcounter{table}{1}
\begin{table*}\caption{The sample stars with spectroscopic [Fe/H] observation. }
\scriptsize
\begin{tabular}{lrrrrrrrr}
\hline\hline
Name & Sp. type & [Fe/H] & [Ca/Fe] & age (Gyr) & $R$ (kpc)& $e$ & $z_{\rm max}$ (kpc)& population \\ \hline  
HD400&	F8IV        &	-0.44&	0.08&	5.39&	8.02&	0.11&	0.03&	thin\\
HD693&	F5V         &	-0.54&	0.03&	5.49&	7.86&	0.09&	0.14&	thin\\
HD739&	F3/F5V      &	-0.27&	0.05&	3.19&	8.33&	0.05&	0.04&	thin\\
HD1581&	F9V         &	-0.29&	-0.19&	5.23&	8.43&	0.19&	0.50&	thin\\
HD1671&	F5III       &	-0.19&	&	1.56&	8.17&	0.02&	0.08&	thin\\
HD2025&	K2V         &	-0.19&	&	&	7.63&	0.14&	0.03&	thin\\
HD2151&	G2IV        &	-0.07&	&	6.28&	7.00&	0.23&	0.29&	thin\\
\hline
\end{tabular}
\normalsize
\label{spec_table}
\end{table*}
\clearpage

\setcounter{table}{1}
\begin{table*}\caption{(continued)}
\scriptsize
\begin{tabular}{lrrrrrrrr}
\hline\hline
Name & Sp. type & [Fe/H] & [Ca/Fe] & age (Gyr) & $R$ (kpc)& $e$ & $z_{\rm max}$ (kpc)& population \\ \hline  
HD20407&	G3V         &	-0.45&	&	3.50&	8.89&	0.10&	0.10&	thin\\
HD20619&	G0          &	-0.21&	&	2.19&	7.31&	0.10&	0.02&	thin\\
HD21197&	K5V         &	0.33&	0.11&	&	7.54&	0.19&	0.19&	thin\\
HD21770&	F4III    &	-0.32&	&	2.75&	7.60&	0.18&	0.04&	thin\\
HD21794&	F7V         &	-0.27&	&	2.12&	7.36&	0.25&	0.18&	thin\\
HD22328&	F5          &	-0.21&	&	2.88&	7.15&	0.17&	0.00&	thin\\
HD22484&	F9V         &	-0.21&	0.03&	7.00&	7.78&	0.04&	0.42&	thin\\
\hline
\end{tabular} 
\normalsize
\end{table*}
\clearpage

\setcounter{table}{1}
\begin{table*}\caption{(continued)}
\scriptsize
\begin{tabular}{lrrrrrrrr}
\hline\hline
Name & Sp. type & [Fe/H] & [Ca/Fe] & age (Gyr) & $R$ (kpc)& $e$ & $z_{\rm max}$ (kpc)& population \\ \hline  
HD66011&	G0IV        &	0.25&	&	2.60&	8.39&	0.05&	0.11&	thin\\
HD66573&	G0          &	-0.57&	0.20&	11.31&	8.71&	0.16&	0.39&	thin\\
HD67228&	G2IV        &	0.18&	-0.01&	5.98&	8.53&	0.14&	0.12&	thin\\
HD67458&	G2V         &	-0.24&	&	3.82&	8.41&	0.22&	0.21&	thin\\
HD67483&	F3V         &	-0.32&	&	1.56&	8.26&	0.07&	0.01&	thin\\
HD68456&	F5V         &	-0.36&	0.08&	3.99&	7.82&	0.10&	0.34&	thin\\
HD69830&	K0V         &	0.06&	0.04&	&	6.59&	0.26&	0.05&	thin\\
\hline
\end{tabular}
\normalsize
\end{table*}

\setcounter{table}{1}
\begin{table*}\caption{(continued)}
\scriptsize
\begin{tabular}{lrrrrrrrr}
\hline\hline
Name & Sp. type & [Fe/H] & [Ca/Fe] & age (Gyr) & $R$ (kpc)& $e$ & $z_{\rm max}$ (kpc)& population \\ \hline  
HD106156&	G8V         &	0.13&	0.08&	4.55&	7.67&	0.20&	0.17&	thin\\
HD107113&	F4V         &	-0.69&	0.09&	5.41&	8.98&	0.17&	0.12&	thin\\
HD107213&	F8Vs        &	-0.04&	-0.01&	3.72&	6.82&	0.18&	0.08&	thin\\
HD108076&	G0V         &	-0.96&	&	6.04&	7.37&	0.29&	0.09&	thin\\
HD108954&	F9V         &	-0.19&	0.01&	3.81&	8.60&	0.08&	0.26&	thin\\
HD109358&	G0V         &	-0.19&	-0.01&	4.05&	8.15&	0.07&	0.09&	thin\\
HD110010&	G0          &	0.18&	-0.01&	6.39&	7.49&	0.07&	0.09&	thin\\
\hline
\end{tabular}
\normalsize
\end{table*}

\setcounter{table}{1}
\begin{table*}\caption{(continued)}
\scriptsize
\begin{tabular}{lrrrrrrrr}
\hline\hline
Name & Sp. type & [Fe/H] & [Ca/Fe] & age (Gyr) & $R$ (kpc)& $e$ & $z_{\rm max}$ (kpc)& population \\ \hline  
HD144172&	F8          &	-0.58&	0.01&	4.23&	7.87&	0.09&	0.03&	thin\\
HD144253&	K3/K4V      &	-0.03&	&	&	8.10&	0.17&	0.23&	thin\\
HD144284&	F8IV-V      &	-0.07&	&	2.73&	7.71&	0.11&	0.14&	thin\\
HD144585&	G5V         &	0.30&	-0.03&	5.43&	7.57&	0.08&	0.28&	thin\\
HD144628&	K3V         &	-0.33&	&	&	7.97&	0.15&	0.32&	thin\\
HD145675&	K0V         &	-0.00&	&	&	8.09&	0.11&	0.05&	thin\\
HD150012&	F5III-IV    &	-0.08&	0.26&	1.87&	7.32&	0.09&	0.06&	thin\\
\hline
\end{tabular}
\normalsize
\end{table*}

\setcounter{table}{1}
\begin{table*}\caption{(continued)}
\scriptsize
\begin{tabular}{lrrrrrrrr}
\hline\hline
Name & Sp. type & [Fe/H] & [Ca/Fe] & age (Gyr) & $R$ (kpc)& $e$ & $z_{\rm max}$ (kpc)& population \\ \hline  
HD182274&	F6V         &	-0.34&	-0.08&	1.73&	7.34&	0.09&	0.05&	thin\\
HD182900&	F6III       &	-0.09&	&	1.99&	7.77&	0.08&	0.13&	thin\\
HD184601&	G0          &	-0.81&	0.22&	9.51&	6.68&	0.27&	0.11&	thin\\
HD184663&	F6IV        &	-0.29&	&	2.29&	8.67&	0.08&	0.10&	thin\\
HD184960&	F7V         &	-0.32&	&	3.86&	8.37&	0.10&	0.08&	thin\\
HD186379&	F8V         &	-0.53&	0.10&	7.25&	7.53&	0.15&	0.49&	thin\\
HD187098&	F3V         &	-0.38&	&	2.84&	7.62&	0.11&	0.09&	thin\\
\hline
\end{tabular}
\normalsize
\end{table*}

\setcounter{table}{1}
\begin{table*}\caption{(continued)}
\scriptsize
\begin{tabular}{lrrrrrrrr}
\hline\hline
Name & Sp. type & [Fe/H] & [Ca/Fe] & age (Gyr) & $R$ (kpc)& $e$ & $z_{\rm max}$ (kpc)& population \\ \hline  
HD216756&	F5II        &	-0.34&	&	3.10&	7.29&	0.10&	0.11&	thin\\
HD217014&	G5V         &	0.17&	-0.05&	6.34&	7.33&	0.09&	0.24&	thin\\
HD217107&	G8IV        &	0.30&	-0.12&	9.47&	7.99&	0.03&	0.17&	thin\\
HD217877&	F8V         &	-0.30&	&	7.43&	7.74&	0.12&	0.01&	thin\\
HD218209&	G6V         &	-0.52&	0.26&	6.15&	6.99&	0.26&	0.09&	thin\\
HD218235&	F6Vs        &	-0.00&	&	3.34&	7.73&	0.12&	0.05&	thin\\
HD218261&	F7V         &	-0.19&	&	4.25&	7.74&	0.08&	0.06&	thin\\
\hline
\end{tabular}
\normalsize
\end{table*}

\setcounter{table}{1}
\begin{table*}\caption{(continued)}
\scriptsize
\begin{tabular}{lrrrrrrrr}
\hline\hline
Name & Sp. type & [Fe/H] & [Ca/Fe] & age (Gyr) & $R$ (kpc)& $e$ & $z_{\rm max}$ (kpc)& population \\ \hline  
HD99109&	K0          &	0.45&	0.00&	14.84&	6.35&	0.26&	0.11&	thick\\
HD103932&	K4V         &	0.16&	-0.28&	&	6.21&	0.29&	0.01&	thick\\
HD108309&	G5IV-V      &	0.22&	-0.01&	8.04&	6.46&	0.29&	0.04&	thick\\
HD108564&	K2V         &	-0.63&	&	&	6.10&	0.34&	0.86&	thick\\
HD110897&	G0V         &	-0.70&	0.08&	6.81&	8.91&	0.14&	1.37&	thick\\
HD114606&	G1V         &	-0.57&	0.21&	11.04&	8.95&	0.45&	1.19&	thick\\
HD114762&	F9V         &	-0.75&	0.21&	8.43&	6.67&	0.33&	0.94&	thick\\
\hline
\end{tabular}
\normalsize
\end{table*}
\clearpage

\setcounter{table}{1}
\begin{table*}\caption{(continued)}
\scriptsize
\begin{tabular}{lrrrrrrrr}
\hline\hline
Name & Sp. type & [Fe/H] & [Ca/Fe] & age (Gyr) & $R$ (kpc)& $e$ & $z_{\rm max}$ (kpc)& population \\ \hline  
HD121004&	G1IV-V      &	-0.73&	0.19&	4.19&	12.15&	0.97&	5.05&	halo\\
HD121849&	G5V         &	-0.28&	&	&	5.51&	0.45&	0.10&	halo\\
HD134113&	F9V         &	-0.83&	0.36&	12.20&	5.77&	0.58&	0.42&	halo\\
HD140283&	sdF3        &	-1.96&	0.21&	11.38&	28.14&	0.99&	4.81&	halo\\
HD148816&	F8V         &	-0.74&	0.33&	8.36&	5.02&	0.90&	2.88&	halo\\
HD155918&	G2V         &	-0.65&	&	10.68&	5.35&	0.50&	1.15&	halo\\
HD160693&	G0V         &	-0.70&	&	9.07&	8.59&	0.73&	2.07&	halo\\
HD176021&	G0/G1III    &	-0.66&	-0.02&	12.00&	5.71&	0.65&	0.30&	halo\\
HD184499&	G0V         &	-0.67&	0.18&	11.73&	4.93&	0.68&	1.34&	halo\\
HD188510&	G5Vwe       &	-1.50&	&	&	6.55&	0.60&	1.18&	halo\\
HD193901&	F7V         &	-1.25&	0.12&	&	337.19&	1.00&	0.13&	halo\\
HD194598&	F7V-VI      &	-1.10&	0.28&	7.66&	4.59&	0.82&	0.41&	halo\\
HD196892&	G2w...      &	-1.00&	0.22&	6.55&	5.25&	0.53&	0.33&	halo\\
HD198245&	sdK0        &	-0.78&	&	10.43&	5.26&	0.52&	0.25&	halo\\
HD201891&	F8V-VI      &	-0.95&	0.18&	8.03&	5.99&	0.54&	0.75&	halo\\
HD204155&	G5          &	-0.80&	0.29&	10.66&	5.32&	0.51&	0.47&	halo\\
HD221830&	F9V         &	-0.46&	0.24&	10.06&	5.69&	0.47&	1.02&	halo\\

\hline
\end{tabular}
\normalsize
\end{table*}

\clearpage
\begin{table*}
\caption{The sample stars with $ubvy$-H$\beta$ photometry.}
\scriptsize
\begin{tabular}{lrrrrrrr}
\hline\hline
Name & Sp. type & [Fe/H] & age (Gyr) & $R$(kpc) & $e$ & $z_{\rm max}$ (kpc) & population \\ \hline  
BD -13 2948&	K0          &	-0.51&	4.52&	9.70&	0.27&	0.18&	thin\\
BD -11 220&	G0          &	-0.51&	5.62&	7.01&	0.34&	0.35&	thin\\
BD 4 415&	K3V         &	-0.41&	&	7.76&	0.45&	0.28&	thin\\
BD 7 385&	K8          &	-0.28&	&	7.59&	0.09&	0.52&	thin\\
BD 9 1627&	K2V         &	-0.08&	7.98&	7.69&	0.33&	0.01&	thin\\
BD 13 374&	K0          &	-0.10&	12.03&	7.30&	0.22&	0.08&	thin\\
BD 16 71&	K3          &	-0.24&	&	6.96&	0.16&	0.09&	thin\\
\hline
\end{tabular} \normalsize
\label{photo_table}
\end{table*}
\normalsize

\setcounter{table}{2}
\begin{table*}\caption{(continued)}
\scriptsize
\begin{tabular}{lrrrrrrr}
\hline\hline
Name & Sp. type & [Fe/H] & age (Gyr) & $R$(kpc) & $e$ & $z_{\rm max}$ (kpc) & population \\ \hline  
HD4247&	F0V         &	-0.40&	1.54&	8.83&	0.10&	0.06&	thin\\
HD4304&	F7III       &	-0.10&	3.60&	7.32&	0.13&	0.09&	thin\\
HD4309&	F8V         &	-0.12&	2.83&	6.76&	0.19&	0.00&	thin\\
HD4975&	G1V         &	-0.11&	5.88&	7.84&	0.04&	0.10&	thin\\
HD5057&	F6V         &	-0.20&	3.61&	8.13&	0.11&	0.25&	thin\\
HD5065&	G0          &	-0.26&	7.29&	7.13&	0.25&	0.59&	thin\\
HD5072&	F7IV        &	-0.47&	8.63&	8.19&	0.03&	0.56&	thin\\
\hline
\end{tabular} \normalsize
\end{table*}
\normalsize
\setcounter{table}{2}
\begin{table*}\caption{(continued)}
\scriptsize
\begin{tabular}{lrrrrrrr}
\hline\hline
Name & Sp. type & [Fe/H] & age (Gyr) & $R$(kpc) & $e$ & $z_{\rm max}$ (kpc) & population \\ \hline  
HD13133&	F5V         &	-0.69&	4.92&	8.47&	0.10&	0.36&	thin\\
HD13201&	F5V         &	-0.43&	4.98&	7.17&	0.12&	0.19&	thin\\
HD13228&	F8          &	-0.32&	4.16&	8.62&	0.15&	0.00&	thin\\
HD13540&	F5          &	-0.63&	2.58&	8.94&	0.11&	0.04&	thin\\
HD13836&	G8V         &	0.01&	&	6.92&	0.18&	0.01&	thin\\
HD13871&	F6IV-V      &	-0.11&	3.16&	7.72&	0.10&	0.08&	thin\\
HD13997&	G5          &	0.10&	9.51&	6.66&	0.20&	0.12&	thin\\
\hline
\end{tabular} \normalsize
\end{table*}
\normalsize

\setcounter{table}{2}
\begin{table*}\caption{(continued)}
\scriptsize
\begin{tabular}{lrrrrrrr}
\hline\hline
Name & Sp. type & [Fe/H] & age (Gyr) & $R$(kpc) & $e$ & $z_{\rm max}$ (kpc) & population \\ \hline  
HD22594&	F3V         &	-0.44&	4.45&	7.45&	0.09&	0.25&	thin\\
HD22596&	F3V         &	-0.50&	4.13&	7.34&	0.09&	0.29&	thin\\
HD23140&	K2          &	-0.14&	&	7.58&	0.10&	0.19&	thin\\
HD23308&	F8V         &	-0.31&	5.13&	7.87&	0.03&	0.37&	thin\\
HD23456&	G1V         &	-0.50&	4.32&	8.83&	0.25&	0.22&	thin\\
HD23609&	F8IV        &	-0.02&	3.88&	8.26&	0.08&	0.02&	thin\\
HD23626&	G0          &	-0.44&	2.98&	8.15&	0.05&	0.04&	thin\\
\hline
\end{tabular} \normalsize
\end{table*}
\normalsize

\setcounter{table}{2}
\begin{table*}\caption{(continued)}
\scriptsize
\begin{tabular}{lrrrrrrr}
\hline\hline
Name & Sp. type & [Fe/H] & age (Gyr) & $R$(kpc) & $e$ & $z_{\rm max}$ (kpc) & population \\ \hline  
HD28483&	F6V         &	-0.11&	2.36&	7.71&	0.10&	0.07&	thin\\
HD28608&	F5          &	-0.15&	2.11&	7.69&	0.10&	0.06&	thin\\
HD28620&	F5          &	-0.62&	3.05&	8.33&	0.06&	0.07&	thin\\
HD28635&	F9V         &	-0.10&	2.18&	7.66&	0.12&	0.04&	thin\\
HD28736&	F5V         &	-0.07&	2.02&	7.64&	0.11&	0.06&	thin\\
HD28911&	F5V         &	-0.16&	2.12&	7.68&	0.09&	0.07&	thin\\
HD29137&	G5V         &	0.17&	8.98&	7.41&	0.34&	0.04&	thin\\
\hline
\end{tabular} \normalsize
\end{table*}
\normalsize

\setcounter{table}{2}
\begin{table*}\caption{(continued)}
\scriptsize
\begin{tabular}{lrrrrrrr}
\hline\hline
Name & Sp. type & [Fe/H] & age (Gyr) & $R$(kpc) & $e$ & $z_{\rm max}$ (kpc) & population \\ \hline  
HD39192&	G0V         &	-0.45&	6.74&	8.56&	0.09&	0.15&	thin\\
HD39427&	G6V         &	-0.31&	4.38&	7.80&	0.15&	0.08&	thin\\
HD39655&	F2V         &	-0.22&	&	7.43&	0.08&	0.08&	thin\\
HD39962&	F2V         &	-0.48&	3.32&	8.43&	0.12&	0.19&	thin\\
HD40125&	F3IV/V      &	-0.10&	2.14&	7.93&	0.07&	0.01&	thin\\
HD40421&	F0          &	-0.11&	2.16&	7.54&	0.07&	0.10&	thin\\
HD40787&	F5          &	-0.23&	2.99&	8.27&	0.10&	0.15&	thin\\
\hline
\end{tabular} \normalsize
\end{table*}
\normalsize

\setcounter{table}{2}
\begin{table*}\caption{(continued)}
\scriptsize
\begin{tabular}{lrrrrrrr}
\hline\hline
Name & Sp. type & [Fe/H] & age (Gyr) & $R$(kpc) & $e$ & $z_{\rm max}$ (kpc) & population \\ \hline  
HD62196&	F0          &	-0.77&	&	8.94&	0.11&	0.20&	thin\\
HD62549&	G1V         &	-0.09&	7.11&	7.78&	0.25&	0.43&	thin\\
HD62613&	G8V         &	-0.13&	&	8.63&	0.07&	0.39&	thin\\
HD63332&	F6V         &	-0.26&	4.07&	8.55&	0.07&	0.01&	thin\\
HD64184&	G5V         &	-0.09&	6.70&	9.30&	0.22&	0.56&	thin\\
HD64685&	F2IV        &	-0.07&	1.45&	7.45&	0.07&	0.06&	thin\\
HD64913&	F8          &	-0.59&	9.77&	7.25&	0.32&	0.30&	thin\\
\hline
\end{tabular} \normalsize
\end{table*}
\normalsize

\setcounter{table}{2}
\begin{table*}\caption{(continued)}
\scriptsize
\begin{tabular}{lrrrrrrr}
\hline\hline
Name & Sp. type & [Fe/H] & age (Gyr) & $R$(kpc) & $e$ & $z_{\rm max}$ (kpc) & population \\ \hline  
HD83509&	F7V         &	-0.63&	5.18&	8.43&	0.05&	0.12&	thin\\
HD83683&	F8          &	-0.37&	3.75&	8.02&	0.10&	0.16&	thin\\
HD84709&	F3V         &	-0.23&	2.12&	7.36&	0.09&	0.07&	thin\\
HD85217&	F6V...      &	-0.21&	3.87&	7.51&	0.09&	0.08&	thin\\
HD85249&	F7V         &	-0.01&	2.18&	8.58&	0.08&	0.05&	thin\\
HD85380&	F8V         &	-0.12&	5.00&	7.58&	0.11&	0.19&	thin\\
HD85683&	F8V         &	-0.07&	3.97&	7.84&	0.08&	0.19&	thin\\
\hline
\end{tabular} \normalsize
\end{table*}
\normalsize

\setcounter{table}{2}
\begin{table*}\caption{(continued)}
\scriptsize
\begin{tabular}{lrrrrrrr}
\hline\hline
Name & Sp. type & [Fe/H] & age (Gyr) & $R$(kpc) & $e$ & $z_{\rm max}$ (kpc) & population \\ \hline  
HD95216&	F5V         &	-0.29&	3.03&	7.84&	0.13&	0.07&	thin\\
HD95363&	F7V         &	-0.36&	4.72&	8.29&	0.04&	0.17&	thin\\
HD95366&	K2          &	-0.06&	&	7.09&	0.29&	0.22&	thin\\
HD95532&	F7/F8V      &	-0.39&	5.14&	8.32&	0.04&	0.40&	thin\\
HD95547&	F8Vm        &	-0.68&	4.67&	9.16&	0.13&	0.52&	thin\\
HD95980&	G5          &	-0.07&	6.37&	7.75&	0.21&	0.03&	thin\\
HD96418&	F8IV        &	-0.22&	4.61&	8.02&	0.00&	0.09&	thin\\
\hline
\end{tabular} \normalsize
\end{table*}
\normalsize

\setcounter{table}{2}
\begin{table*}\caption{(continued)}
\scriptsize
\begin{tabular}{lrrrrrrr}
\hline\hline
Name & Sp. type & [Fe/H] & age (Gyr) & $R$(kpc) & $e$ & $z_{\rm max}$ (kpc) & population \\ \hline  
HD104755&	F5          &	-0.33&	2.76&	6.94&	0.16&	0.07&	thin\\
HD104868&	F6V         &	-0.35&	2.08&	8.01&	0.09&	0.16&	thin\\
HD104904&	F6V         &	-0.46&	1.83&	8.82&	0.10&	0.26&	thin\\
HD104982&	G5V         &	-0.14&	2.01&	7.98&	0.22&	0.54&	thin\\
HD104988&	G8V         &	-0.04&	&	7.31&	0.23&	0.10&	thin\\
HD104999&	F2          &	-0.18&	2.87&	7.04&	0.14&	0.18&	thin\\
HD105087&	K0          &	-0.41&	&	8.15&	0.34&	0.43&	thin\\
\hline
\end{tabular} \normalsize
\end{table*}
\normalsize

\clearpage
\setcounter{table}{2}
\begin{table*}\caption{(continued)}
\scriptsize
\begin{tabular}{lrrrrrrr}
\hline\hline
Name & Sp. type & [Fe/H] & age (Gyr) & $R$(kpc) & $e$ & $z_{\rm max}$ (kpc) & population \\ \hline  
HD110477&	F6IV        &	-0.18&	2.27&	7.20&	0.19&	0.13&	thin\\
HD110524&	F4V         &	-0.29&	2.10&	7.59&	0.07&	0.05&	thin\\
HD110619&	G5V         &	-0.39&	&	7.53&	0.21&	0.38&	thin\\
HD110833&	K3V         &	-0.20&	&	7.55&	0.07&	0.21&	thin\\
HD110949&	F8V         &	-0.30&	2.22&	8.40&	0.05&	0.29&	thin\\
HD111238&	F8V         &	-0.34&	4.51&	7.82&	0.04&	0.02&	thin\\
HD111347&	F7V         &	-0.30&	4.40&	8.03&	0.05&	0.04&	thin\\
\hline
\end{tabular} \normalsize
\end{table*}
\normalsize

\setcounter{table}{2}
\begin{table*}\caption{(continued)}
\scriptsize
\begin{tabular}{lrrrrrrr}
\hline\hline
Name & Sp. type & [Fe/H] & age (Gyr) & $R$(kpc) & $e$ & $z_{\rm max}$ (kpc) & population \\ \hline  
HD120530&	F5          &	-0.23&	2.03&	7.94&	0.01&	0.01&	thin\\
HD120544&	F6IV/V      &	-0.10&	1.42&	8.35&	0.06&	0.02&	thin\\
HD120623&	F0/F2IV     &	-0.32&	1.30&	8.04&	0.07&	0.09&	thin\\
HD120672&	F6V         &	-0.14&	2.74&	7.16&	0.12&	0.16&	thin\\
HD120915&	F5          &	-0.01&	3.56&	8.22&	0.06&	0.22&	thin\\
HD121105&	F5          &	-0.27&	3.74&	7.25&	0.11&	0.22&	thin\\
HD121249&	K0          &	-0.10&	&	8.26&	0.25&	0.09&	thin\\
\hline
\end{tabular} \normalsize
\end{table*}
\normalsize

\setcounter{table}{2}
\begin{table*}\caption{(continued)}
\scriptsize
\begin{tabular}{lrrrrrrr}
\hline\hline
Name & Sp. type & [Fe/H] & age (Gyr) & $R$(kpc) & $e$ & $z_{\rm max}$ (kpc) & population \\ \hline  
HD138525&	F6III       &	-0.19&	2.42&	7.32&	0.09&	0.39&	thin\\
HD139389&	F5V:        &	-0.41&	2.49&	8.22&	0.06&	0.16&	thin\\
HD139590&	G0V         &	-0.13&	3.86&	6.70&	0.21&	0.12&	thin\\
HD139664&	F5IV-V      &	-0.27&	2.20&	7.59&	0.06&	0.04&	thin\\
HD140812&	F5V         &	-0.49&	3.68&	9.18&	0.15&	0.22&	thin\\
HD141247&	F9V         &	-0.19&	5.59&	7.10&	0.13&	0.23&	thin\\
HD141465&	F3V         &	-0.31&	&	8.34&	0.06&	0.05&	thin\\
\hline
\end{tabular} \normalsize
\end{table*}
\normalsize

\setcounter{table}{2}
\begin{table*}\caption{(continued)}
\scriptsize
\begin{tabular}{lrrrrrrr}
\hline\hline
Name & Sp. type & [Fe/H] & age (Gyr) & $R$(kpc) & $e$ & $z_{\rm max}$ (kpc) & population \\ \hline  
HD158633&	K0V         &	-0.43&	&	6.76&	0.19&	0.12&	thin\\
HD159517&	F4V         &	-0.15&	1.78&	8.07&	0.13&	0.09&	thin\\
HD159868&	G5V         &	0.04&	6.87&	6.59&	0.23&	0.43&	thin\\
HD160043&	F5V         &	-0.44&	3.37&	7.04&	0.14&	0.17&	thin\\
HD160346&	K3V         &	-0.23&	&	8.32&	0.09&	0.20&	thin\\
HD160482&	F5V         &	-0.27&	3.92&	6.79&	0.19&	0.01&	thin\\
HD160487&	F5          &	-0.44&	2.52&	7.39&	0.21&	0.10&	thin\\
\hline
\end{tabular} \normalsize
\end{table*}
\normalsize

\setcounter{table}{2}
\begin{table*}\caption{(continued)}
\scriptsize
\begin{tabular}{lrrrrrrr}
\hline\hline
Name & Sp. type & [Fe/H] & age (Gyr) & $R$(kpc) & $e$ & $z_{\rm max}$ (kpc) & population \\ \hline  
HD181144&	F8IV-V      &	-0.30&	4.99&	8.05&	0.03&	0.11&	thin\\
HD181428&	G1V         &	-0.13&	5.32&	8.84&	0.14&	0.43&	thin\\
HD181544&	G1V         &	-0.16&	5.16&	9.11&	0.15&	0.51&	thin\\
HD182488&	G8V         &	0.03&	10.53&	7.80&	0.04&	0.04&	thin\\
HD182901&	F5III       &	-0.13&	2.15&	7.72&	0.11&	0.15&	thin\\
HD182926&	F5V         &	-0.33&	2.64&	7.99&	0.04&	0.03&	thin\\
HD183028&	F5V         &	-0.24&	2.52&	8.17&	0.05&	0.05&	thin\\
\hline
\end{tabular} \normalsize
\end{table*}
\normalsize

\setcounter{table}{2}
\begin{table*}\caption{(continued)}
\scriptsize
\begin{tabular}{lrrrrrrr}
\hline\hline
Name & Sp. type & [Fe/H] & age (Gyr) & $R$(kpc) & $e$ & $z_{\rm max}$ (kpc) & population \\ \hline  
HD198585&	G0          &	-0.62&	2.93&	9.33&	0.14&	0.30&	thin\\
HD198802&	G1V         &	-0.02&	4.25&	8.54&	0.06&	0.04&	thin\\
HD199190&	G5IV        &	0.11&	6.32&	7.86&	0.26&	0.23&	thin\\
HD199260&	F7V         &	-0.26&	3.46&	7.90&	0.02&	0.08&	thin\\
HD199373&	F5V         &	-0.41&	4.67&	6.79&	0.19&	0.08&	thin\\
HD200334&	G3IV        &	0.04&	4.28&	7.67&	0.08&	0.07&	thin\\
HD200433&	F5          &	-0.33&	2.31&	8.63&	0.09&	0.08&	thin\\
\hline
\end{tabular} \normalsize
\end{table*}
\normalsize

\setcounter{table}{2}
\begin{table*}\caption{(continued)}
\scriptsize
\begin{tabular}{lrrrrrrr}
\hline\hline
Name & Sp. type & [Fe/H] & age (Gyr) & $R$(kpc) & $e$ & $z_{\rm max}$ (kpc) & population \\ \hline  
HD212858&	F8          &	-0.51&	9.42&	9.73&	0.24&	0.18&	thin\\
HD213429&	F7V         &	-0.16&	5.40&	7.57&	0.06&	0.01&	thin\\
HD213646&	K2          &	-0.19&	19.89&	8.90&	0.32&	0.12&	thin\\
HD213845&	F7V         &	-0.13&	2.33&	7.56&	0.06&	0.07&	thin\\
HD213928&	F7/F8V      &	-0.56&	5.98&	7.87&	0.11&	0.15&	thin\\
HD213996&	K0          &	-0.03&	&	7.69&	0.32&	0.14&	thin\\
HD214094&	F6V         &	-0.26&	4.45&	8.08&	0.22&	0.18&	thin\\
\hline
\end{tabular} \normalsize
\end{table*}
\normalsize

\setcounter{table}{2}
\begin{table*}\caption{(continued)}
\scriptsize
\begin{tabular}{lrrrrrrr}
\hline\hline
Name & Sp. type & [Fe/H] & age (Gyr) & $R$(kpc) & $e$ & $z_{\rm max}$ (kpc) & population \\ \hline  
HD224839&	F8V         &	-0.33&	9.27&	7.51&	0.07&	0.07&	thin\\
HD225045&	F6/F7V      &	-0.05&	2.46&	8.52&	0.10&	0.09&	thin\\
HD225233&	F2V         &	-0.20&	2.65&	7.63&	0.06&	0.13&	thin\\
HD231825&	G5          &	-0.00&	&	7.26&	0.19&	0.08&	thin\\
HD275458&	K0          &	-0.19&	10.13&	7.13&	0.16&	0.34&	thin\\
HD284414&	K2V         &	-0.20&	&	7.51&	0.12&	0.10&	thin\\
HD285690&	K3V         &	-0.18&	&	7.63&	0.12&	0.08&	thin\\
\hline
\end{tabular} \normalsize
\end{table*}
\normalsize

\setcounter{table}{2}
\begin{table*}\caption{(continued)}
\scriptsize
\begin{tabular}{lrrrrrrr}
\hline\hline
Name & Sp. type & [Fe/H] & age (Gyr) & $R$(kpc) & $e$ & $z_{\rm max}$ (kpc) & population \\ \hline  
HD31501&	G8V         &	-0.19&	1.78&	6.44&	0.30&	0.73&	thick\\
HD36283&	G5V         &	-0.25&	4.89&	6.18&	0.30&	0.82&	thick\\
HD37739&	F5          &	-0.49&	4.78&	6.21&	0.39&	0.50&	thick\\
HD38014&	K1V         &	-0.14&	12.43&	5.72&	0.40&	0.13&	thick\\
HD38510&	F5/F6V      &	-0.78&	9.15&	7.90&	0.53&	1.16&	thick\\
HD40040&	G0          &	-0.30&	10.29&	6.35&	0.33&	0.07&	thick\\
HD42160&	G2V         &	-0.15&	&	6.42&	0.25&	0.25&	thick\\
\hline
\end{tabular} \normalsize
\end{table*}
\normalsize
\clearpage

\setcounter{table}{2}
\begin{table*}\caption{(continued)}
\scriptsize
\begin{tabular}{lrrrrrrr}
\hline\hline
Name & Sp. type & [Fe/H] & age (Gyr) & $R$(kpc) & $e$ & $z_{\rm max}$ (kpc) & population \\ \hline  
HD184855&	G0          &	-0.32&	8.93&	8.69&	0.29&	0.61&	thick\\
HD187645&	G5          &	-0.26&	6.57&	6.48&	0.36&	0.21&	thick\\
HD190649&	G5V         &	-0.49&	10.83&	5.93&	0.38&	0.35&	thick\\
HD191069&	G5V         &	-0.07&	8.11&	5.93&	0.36&	0.98&	thick\\
HD196800&	G1/G2V      &	0.07&	4.94&	6.40&	0.31&	0.04&	thick\\
HD197484&	G2V         &	-0.52&	9.75&	6.50&	0.26&	0.05&	thick\\
HD197666&	G0          &	-0.73&	4.91&	7.39&	0.13&	0.61&	thick\\
\hline
\end{tabular} \normalsize  
\end{table*} 
\normalsize
\clearpage

\bibliographystyle{aa}
\bibliography{refs}

\end{document}